\documentclass[fleqn,usenatbib]{mnras}

\usepackage[T1]{fontenc}

\DeclareRobustCommand{\VAN}[3]{#2}
\let\VANthebibliography\thebibliography
\def\thebibliography{\DeclareRobustCommand{\VAN}[3]{##3}\VANthebibliography}

\usepackage{graphicx}	
\usepackage{amsmath}	
\usepackage{amssymb}	
\usepackage{color,xcolor}
\definecolor{orange}{HTML}{FF7F00}
\usepackage{newtxtext,newtxmath}

\title{Recovering the Wedge Modes Lost to 21-cm Foregrounds}

\author[Gagnon-Hartman et al.]{
    Samuel Gagnon-Hartman$^{1,2}$\thanks{samuel.gagnon-hartman@mail.mcgill.ca},
    Yue Cui$^{1,3}$,
    Adrian Liu$^{1}$,
    Siamak Ravanbakhsh$^{4,5}$
\\
$^{1}$Department of Physics and McGill Space Institute, McGill University, Montreal, QC, Canada H3A 2T8 \\
$^{2}$Department of Physics and Astronomy, Bishop's University, 2600 College Street, Sherbrooke J1M 1Z7, Canada \\
$^{3}$University of Electronic Science and Technology of China, 2006 Xiyuan Ave., West High-tech Zone, Chengdu, Sichuan, China\\
$^{4}$School of Computer Science, McGill University, 845 Sherbrooke Street, Montreal H3A 0G4, Canada \\
$^{5}$MILA - Quebec AI Institute, 6666 St Urbain St, Montreal, Quebec H2S 3H1, Canada
}

\date{Accepted XXX. Received YYY; in original form ZZZ}

\pubyear{2020}

\begin{document}
\label{firstpage}
\pagerange{\pageref{firstpage}--\pageref{lastpage}}
\maketitle

\begin{abstract}
\noindent One of the critical challenges facing imaging studies of the 21-cm signal at the Epoch of Reionization (EoR) is the separation of astrophysical foreground contamination. These foregrounds are known to lie in a wedge-shaped region of $(k_{\perp},k_{\parallel})$ Fourier space. Removing these Fourier modes excises the foregrounds at grave expense to image fidelity, since the cosmological information at these modes is also removed by the wedge filter. However, the 21-cm EoR signal is non-Gaussian, meaning that the lost wedge modes are correlated to the surviving modes by some covariance matrix. We have developed a machine learning-based method which exploits this information to identify ionized regions within a wedge-filtered image. Our method reliably identifies the largest ionized regions and can reconstruct their shape, size, and location within an image. We further demonstrate that our method remains viable when instrumental effects are accounted for, using the Hydrogen Epoch of Reionization Array and the Square Kilometre Array as fiducial instruments. The ability to recover spatial information from wedge-filtered images unlocks the potential for imaging studies using current- and next-generation instruments without relying on detailed models of the astrophysical foregrounds themselves.
\end{abstract}




\section{Introduction}

\begin{figure}
	\includegraphics[width=\columnwidth,trim=2cm 1.5cm 2cm 0cm,clip]{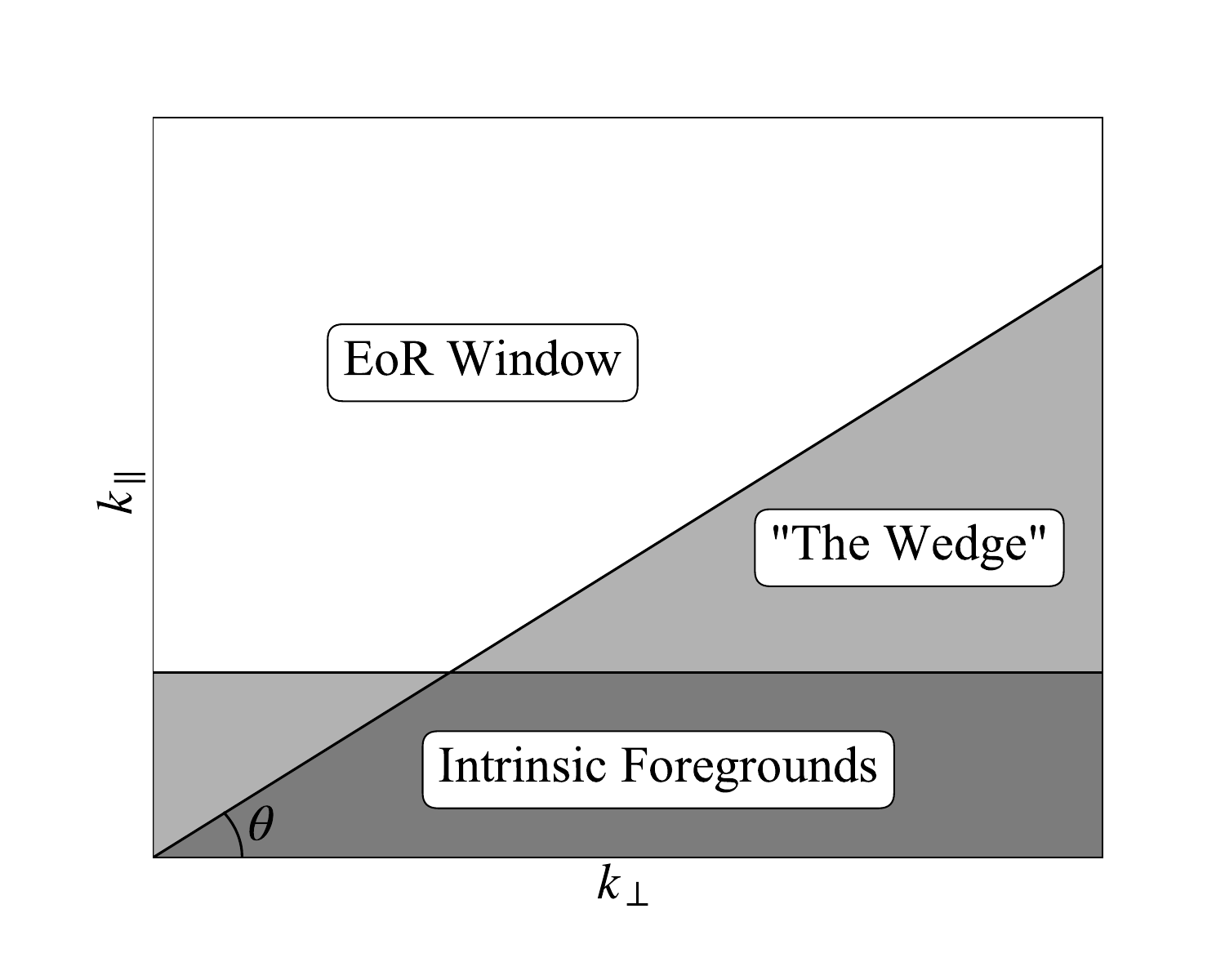}
	\caption{A schematic depicting the footprint of foreground contamination in cylindrical $k$-space. Intrinsic foregrounds uniformly overwhelm low $k_{\parallel}$ modes, while mode-mixing contaminates higher $k_{\parallel}$ modes in a region called the wedge. The region left untouched by foregrounds is referred to as the Epoch of Reionization (EoR) window. $\theta$ designates the ``wedge angle", which characterizes the wedge's footprint and is equivalent to $\mathrm{arctan}(C)$ with reference to Equation \ref{eq:wedge}.}
	\label{fig:eor-window}
\end{figure}

The highly redshifted 21-cm line is becoming recognized as a promising probe of the high-redshift universe, with the potential to use neutral hydrogen as a tracer to map out volumes extending from redshift $z\sim 0$ through the Epoch of Reionization (EoR), Cosmic Dawn, and beyond (for reviews of the field, see \citealt{Furlanetto2006,Morales2010,Pritchard2012,LiuShaw2020}). To be successful, 21-cm experiments must be able to separate the neutral hydrogen signal from bright galactic and extragalactic foregrounds, as these can be brighter than the neutral hydrogen signal by many orders of magnitude (see, e.g., \citealt{Santos2005,GSM,Zheng:2017}).
Most analysis techniques used for removing foregrounds focus on using the spectral smoothness of the foreground emission to distinguish it from the underlying cosmological signal. Numerous techniques have been proposed to remove foregrounds from 21-cm data based on this principle (e.g. \citealt{Morales2006,Wang2006,Bowman2009,Liu2009,Liu2011,Parsons2012,Chapman2012,Chapman2013,Dillon2013,Wolz2017,Carucci2020}).
Studies of the interaction between an interferometer and foreground emission have demonstrated that smooth-spectrum foregrounds occupy an anisotropic wedge-shaped region of Fourier space,
leaving only a small window of Fourier space where the 21-cm signal may be cleanly observed (e.g. \citealt{Parsons2012,Datta2010,Vedantham2012,Morales2012,Trott2012,Thyagarajan2013,Hazelton2013,Liu:2014a,Liu:2014b}). This is illustrated in Figure~\ref{fig:eor-window}, where $k_\perp$ refers to spatial wavenumbers perpendicular to the line of sight of one's observations and $k_\parallel$ to spatial wavenumbers parallel to the line of sight. Up to some proportionality factors, the former is the Fourier dual to angles on the sky, while the latter is the Fourier dual to frequency since the observed redshift of the $21$-cm emission can be mapped to radial distance. 

\begin{figure*}
	\includegraphics[width=\textwidth]{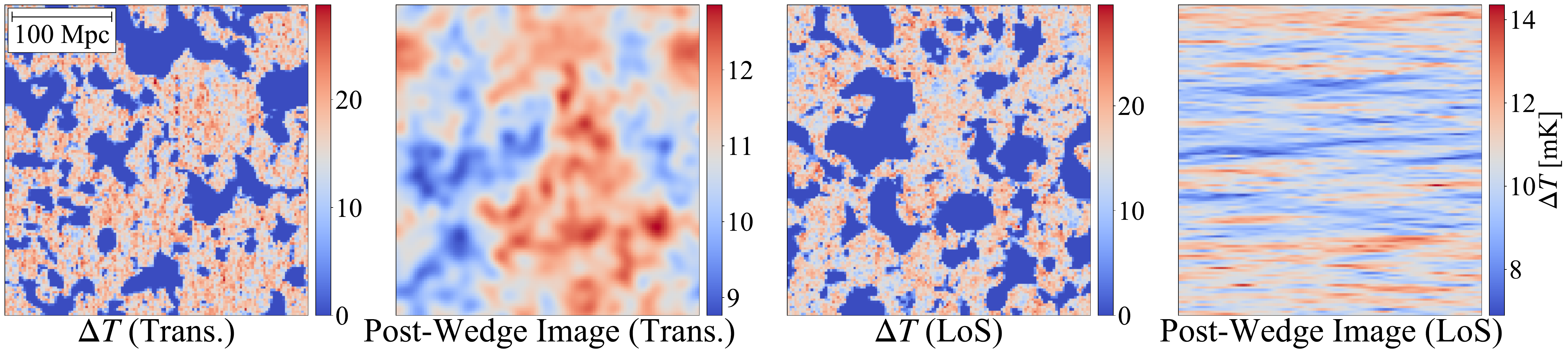}
	\caption{A 21cmFAST 21-cm brightness temperature anisotropy map before and after replacing all Fourier modes within outside the EoR window with null vectors. In the left pair of images, box axes are transverse to the line of sight direction, while in the right pair of images the vertical axis is the line of sight direction. Localized structures are lost in both the transverse and line of sight dimensions.}
	\label{fig:wedge_demo}
\end{figure*}

A clean measurement of the cosmological 21-cm signal can therefore in principle be made by probing only the regions of Fourier space which do not lie within the wedge. \cite{Pober2014a} demonstrates that for statistical quantities like the power spectrum of spatial fluctuations, current instruments such as the Hydrogen Epoch of Reionization Array (\citealt{DeBoer:2017}; HERA) can in principle make high signal-to-noise measurements with this type of ``foreground avoidance" strategy. However, this approach has some drawbacks. First, the cosmological signal strength peaks on large scales (or equivalently, on modes with small spatial wavenumber, $k$), meaning that $k$ modes lying within the wedge may have a significantly higher signal-to-noise ratio than those lying in uncontaminated regions. Foreground avoidance therefore results in a potential significant reduction in the overall signal-to-noise of one's measurement. Second, while statistical measurements like the power spectrum can leverage the statistical isotropy of our Universe in a foreground avoidance scheme, this is not an avenue that is available to imaging experiments, which need to retain full realization-specific information on individual Fourier modes. Said differently, eliminating Fourier modes within the wedge is equivalent to filtering the data in a rather strange way, where the data is put through a high-pass filter in the spectral direction with finer angular scales (high $k_\perp$) being subject to a more aggressive filter. The resulting images therefore become extremely difficult to interpret. This can be seen in Figure~\ref{fig:wedge_demo}, where we show the effect of a foreground wedge filter on noiseless example images of $21$-cm emission during the EoR. Two effects are immediately apparent. The first is that the map is no longer statistically isotropic. The second is that the locations of ionized bubbles (those with zero $21$-cm brightness temperature) around first-generation galaxies are distorted beyond recognition. It is not simply the case that the wedge-filtered maps are slightly blurred versions of the original maps; the morphologies are completely different \citep{Beardsley2015}.

To do EoR science using $21$-cm \emph{images}, one should therefore go beyond foreground avoidance and actually perform foreground subtraction.\footnote{An alternative approach is to forward model the distortions of wedge-filtered images and to make probabilistic statements regarding the true images, as was explored in \citet{Beardsley2015}.} Numerous techniques have been proposed for this (see \citealt{LiuShaw2020} for a summary). Many of these techniques involve the explicit modelling of foreground emission or parameterized fits (whether based on preset templates or empirical ones). Thus far, neither technique has demonstrated that the foreground emission in an actual observation can be removed to the thermal noise level of the instruments. Recently, machine learning-based foreground removal ideas have been explored in the literature. For example, \citet{li2019} trained a convolutional denoising autoencoder to model and remove foreground emission as seen through an instrumental beam pattern, outputting the underlying EoR signal. \citet{Makinen2020} consider hypothetical single-dish $21$-cm observations of the post-reionization neutral hydrogen signal and use a U-Net to improve foreground cleaning following a more traditional principal-component-based foreground removal step.

In this paper, we build on \citet{li2019}, \citet{Makinen2020}, and \citet{Villanueva_Domingo_2021} to propose a U-Net-based deep learning algorithm to recover Fourier modes that are ignored or nulled out by a foreground avoidance scheme. The U-Net architecture adopted in this study is similar to the one presented in \citet{Villanueva_Domingo_2021}, but with modifications to accommodate our 3D data set and to improve performance. In our study, we are asking more of our network than \citet{li2019} did in theirs, because we are attempting to recover Fourier modes after a more aggressive cut; in \citet{li2019}, only the first few $k_\parallel$ modes (corresponding to the ``Intrinsic Foregrounds" portion of Figure~\ref{fig:eor-window}) were excised from the data, whereas we remove the entire foreground wedge and have our network reconstruct the cosmological signal there from the non-excised modes. The initial principal component pre-processing in \citet{Makinen2020} will in principle touch a broad range of Fourier modes. This occurs because systematics tend to proliferate across many Fourier modes \citep{SwitzerLiu2014}. The flip side of this, however, is that the removal of a set of principal components will in general not entirely zero out any Fourier modes. Our work builds on this by considering the recovery of cosmological Fourier modes after a more drastic excision: the relevant Fourier modes are zeroed out completely, and because we are dealing with the instrumentally more complicated case of an interferometer (rather than a single dish), we conservatively excise all Fourier modes within the wedge. Our work assumes a fiducial set of astrophysical and cosmological parameters, and its generalization to a wider range of
parameters still needs to be assessed. Our choice of parametrization was made for ease of comparison to other works (e.g. \citealt{Pober2014a,Gillet_2019}).

After removing the Fourier modes in the foreground wedge, it is unclear \emph{a priori} whether there remains enough information to recover the cosmological portion of the excised modes from the rest of the dataset. If the EoR signal were Gaussian-distributed and obeyed stationary statistics, we would immediately know that this is impossible, since the Fourier modes would then be uncorrelated. However, during the EoR there are significant non-Gaussian correlations between Fourier modes \citep{Shimabukuro_2016,Majumdar_2018,Watkinson_2018,Hutter_2019,Gorce_2019}. This in principle allows a reconstruction of modes that are lost in the foreground suppression (or subtraction) process, and indeed, this is the idea of proposed tidal reconstruction schemes for post-reionization $21$-cm experiments \citep{Zhu2018fgmoderecovery,Li2018kSZtidal,GokselKaracayli2019}. Unfortunately, our relative ignorance of the relevant astrophysics of the EoR makes such a reconstruction difficult to formulate using traditional cosmological techniques. It is for this reason that we turn to a machine-learning-based approach.

In what follows, we will demonstrate with our U-net that there \emph{is} in fact enough information to recover reasonable images of the EoR after completely removing Fourier modes in the foreground wedge. We will focus on an imaging application of $21$-cm maps: the identification of ionized bubbles during the EoR. We will demonstrate that a machine learning approach enables a reliable identification of the largest ionized bubbles, even with current-generation experiments. The rest of the paper is structured as follows. Section \ref{sec:wedge} reviews the phenomenology of the wedge and establishes notation. Section \ref{sec:data} describes the data preparation procedure and the architecture of the Convolutional Neural Network (CNN) that we use. Section \ref{sec:results} includes a description of the five trainings run using the network and their results.

\section{The Foreground Wedge}
\label{sec:wedge}

In this section, we briefly review the foreground wedge. For a more in-depth summary and derivations, see \citet{LiuShaw2020} and references therein. For a review of some alternative foreground removal techniques, see \citet{Hothi_2020,Cunnington2020}.

At the relevant frequencies, astrophysical foregrounds such as Galactic synchrotron emission overwhelm the EoR signal by $4$ to $5$ orders of magnitude, making spatial mapping of the EoR signal impossible without some means of foreground avoidance, removal, or subtraction. Since the foreground elements which contaminate the high-redshift 21-cm signal are expected to be spectrally smooth, only the lowest $k_{\parallel}$ modes (i.e., modes along the line of sight or frequency direction) should be intrinsically affected. However, the frequency dependence inherent to an interferometer's response causes what is referred to as ``mode-mixing", whereby contamination leaks into higher $k_\parallel$ modes. This effect is most pronounced for longer baselines of an interferometer (which probe high $k_{\perp}$ angular Fourier modes), since these baselines have finer fringe patterns that dilate or contract more quickly with changing frequency. The proliferation of foregrounds to a broader range of Fourier modes reduces the available Fourier space over which cosmological measurements can be performed.

Fortunately, the physics of mode-mixing predicts that this proliferation is limited to a well-defined wedge-shaped region of $k_{\perp}$-$k_{\parallel}$ space \citep{Datta2010,Pober2014a,Dillon2014,Liu:2014a,Pober2014b}, illustrated schematically in Figure \ref{fig:eor-window}. Mathematically, the boundary of the wedge is given by
\begin{equation}
	k_{\parallel}=k_{\perp} \left(\textrm{sin}\theta_{\textrm{FoV}}\frac{D_{\mathrm{M}}(z)E(z)}{D_{\mathrm{H}}(1+z)}\right)\equiv   k_{\perp} \tan \psi
	\label{eq:wedge},
\end{equation}
where $\theta_{\mathrm{FoV}}$ is the angular radius of the field of view, $D_{\mathrm{H}}\equiv c/H_0$, $H_0$ is the Hubble parameter, $E(z)\equiv \sqrt{\Omega_\mathrm{m}(1+z)^3+\Omega_\Lambda}$, $\Omega_\mathrm{m}$ is the normalized matter density, $\Omega_\Lambda$ is the normalized dark energy density, and $D_{\mathrm{M}}(z)$ is the transverse co-moving distance \citep{Hogg1999}. We have additionally defined $\psi$ to be the angle that the wedge makes with the $k_\perp$ axis. There is some uncertainty as to precisely what this angle ought to be, because there is a lack of consensus as to what value of $\theta_{\mathrm{FoV}}$ should be inserted into the expression. A pessimistic assumption might be to set $\theta_{\mathrm{FoV}}$ to be $90^\circ$. This corresponds to the horizon, which may be a realistic choice since antenna beam patterns do not generally have sharp cutoffs, and even low-level sidelobes can pick up on bright foreground emission very far away from zenith. More optimistic forecasts in the literature have assumed $\theta_{\mathrm{FoV}}< 90^\circ$, reflecting the community's aspiration that some combination of beam control and foreground subtraction may be able to reduce the bleed of foregrounds in Fourier space.

\begin{figure}
	\includegraphics[width=\columnwidth]{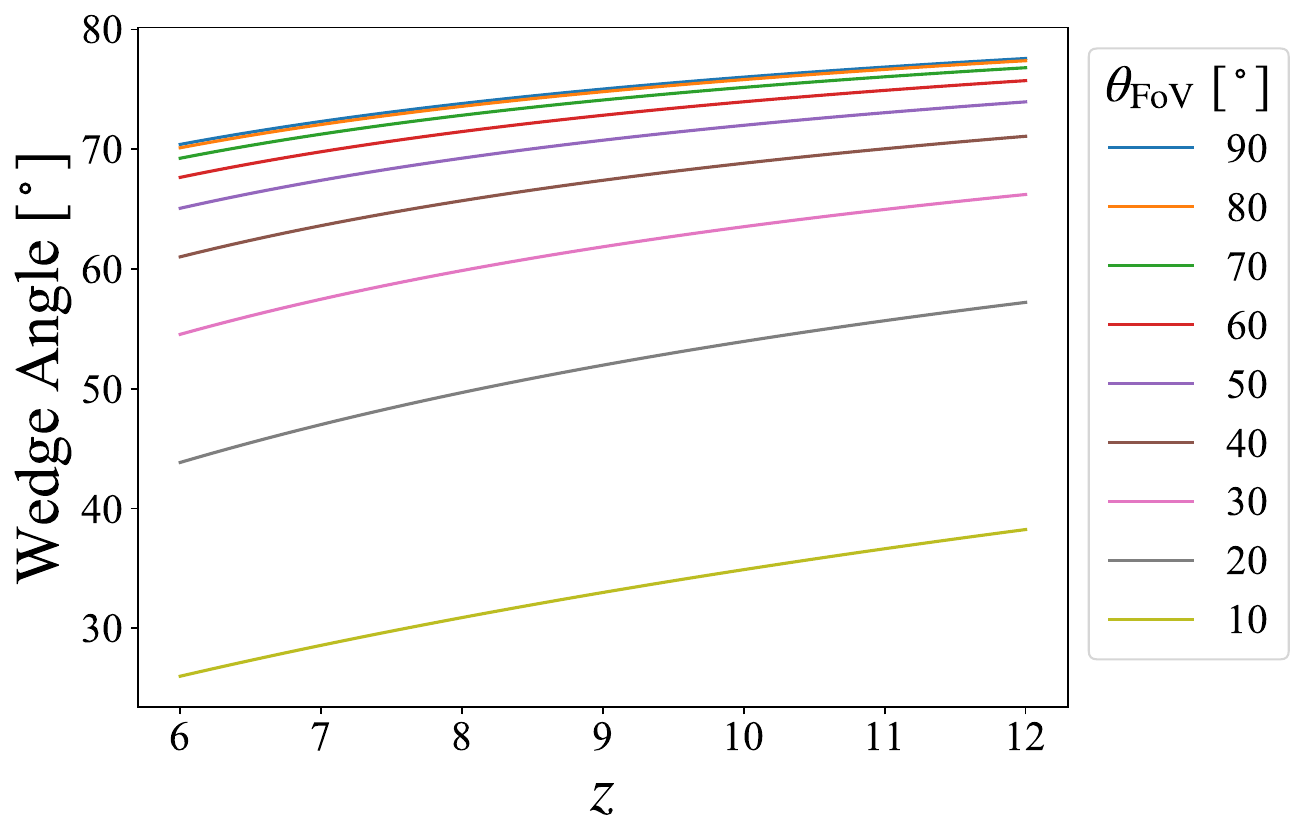}
	\caption{The internal angle of the wedge as a
	function of redshift for various $\theta_{\mathrm{FoV}}$. This work assumes a wedge angle of 75$^\circ$, which is consistent with the most pessimistic possible wedge at $z=8.5$, the highest redshift considered in this study.}
	\label{fig:angles}
\end{figure}

In Figure~\ref{fig:angles} we illustrate how $\psi$ scales with $\theta_{\mathrm{FoV}}$ and redshift $z$. In this paper, we conservatively zero out Fourier modes lying below $\psi = 75^\circ$. This roughly corresponds to the most pessimistic case of a horizon wedge at $z = 8.5$, the highest redshift considered in this study. Such a filter was how Figure~\ref{fig:wedge_demo} was produced, demonstrating that a substantial amount of information is lost. Filtered images like those will be the starting point for our information recovery, and in Section~\ref{sec:data} we go into more detail about our data preparation before showcasing our results in Section~\ref{sec:results}.
%
%
%
%
%

\section{Data Preparation and Network Structure}
\label{sec:data}

\subsection{Forward Simulation}
\label{sec:ForwardSim}
\begin{figure*}
	\includegraphics[width=6 in]{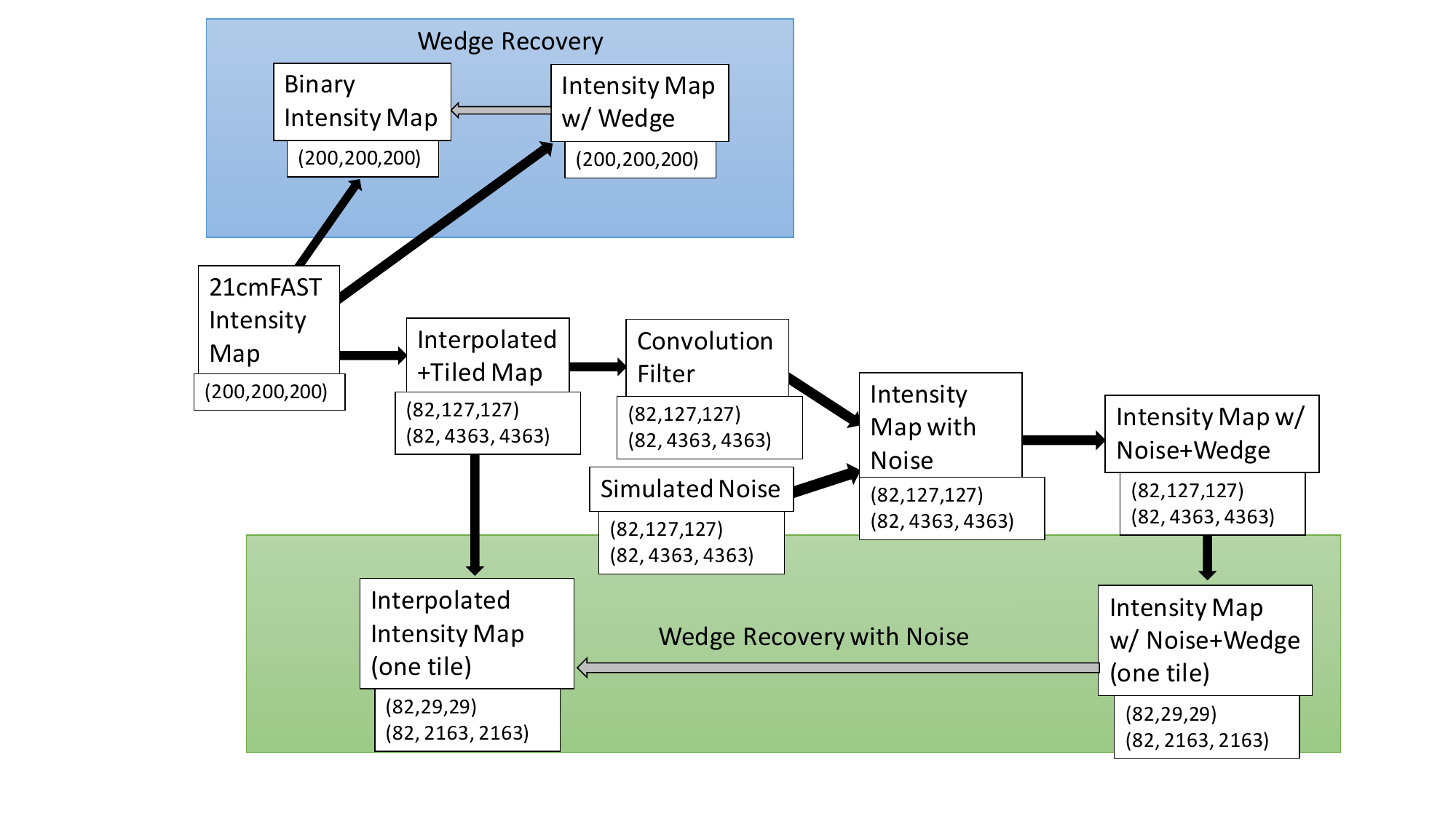}
	\caption{The preprocessing routine used for data preparation. The size of an individual map at each step is given beneath the name of the step. 
	In boxes where two sizes are given, the top corresponds to HERA noise preparation while the bottom corresponds to SKA noise preparation.
	The network used in this paper was tested on removing the wedge from maps where 
	instrumental noise is present and from maps without instrumental noise. The noise-enriched intensity maps are first convolved with the kernel
	inherent to the instrument of interest and then the thermal noise inherent to the instrument is added to the result. The wedge filter
	used in both pathways Fourier transforms the map, sends all vectors within the wedge region to zero, and then inverse Fourier transforms back
	to real space to produce the wedge-affected image.}
	\label{fig:preprocess_flowchart}
\end{figure*}

\begin{figure}
	\includegraphics[width=\columnwidth]{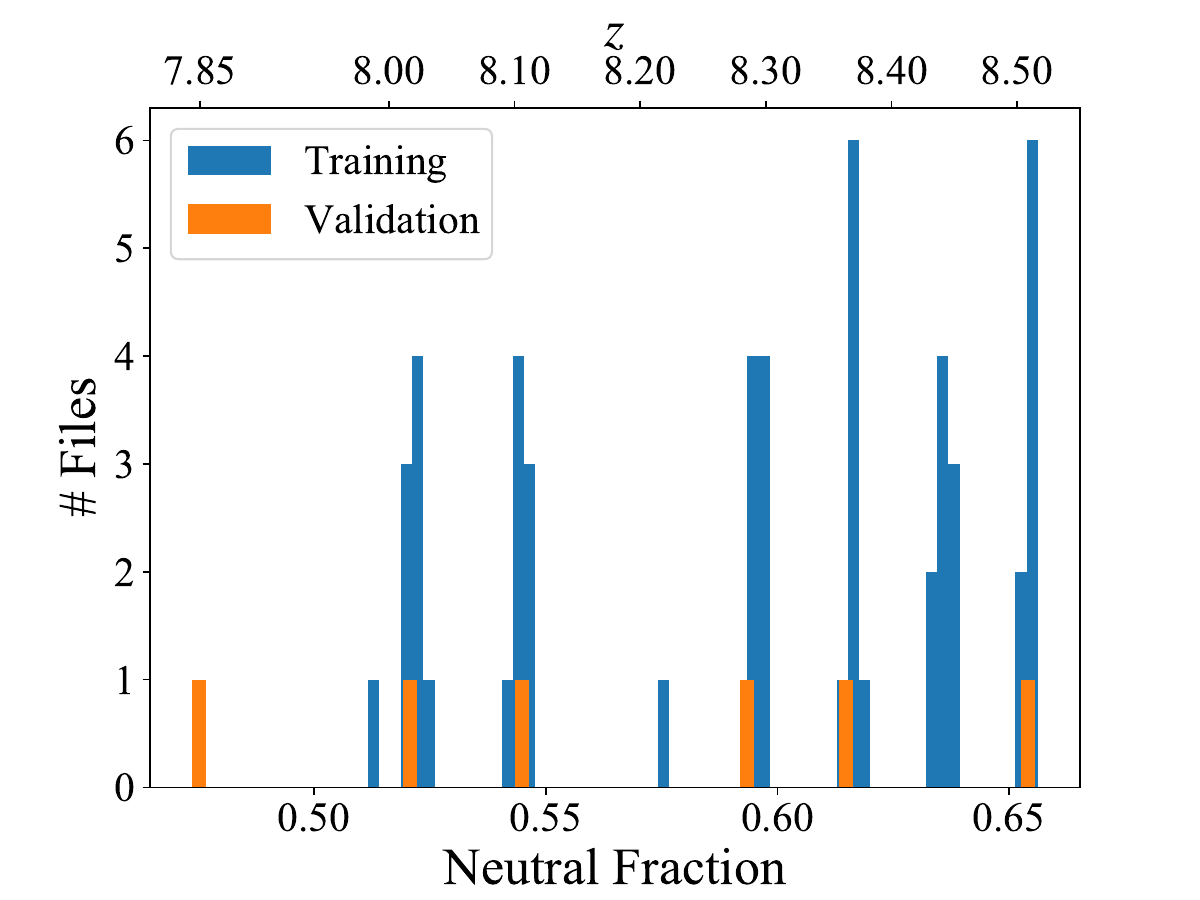}
	\caption{The distribution of training and validation files across neutral fractions. Files are clustered around particular neutral fractions indicative of the 
	redshift from which they were drawn. One validation file was selected from a neutral fraction lower than those represented in the training set.}
	\label{fig:file-dist}
\end{figure}

The data used to represent the ``clean'' 21-cm signal are produced using \texttt{21cmFAST}, a semi-numerical simulation of the highly redshifted 21-cm signal \citep{Mesinger2010}. We chose \texttt{21cmFAST} because its relatively quick speed enables the construction of a sufficient amount of data upon which to train a neural network. The relevant outputs from the code are maps of the $21$-cm brightness temperature field, evaluated at fixed snapshots in redshift. In other words, we do not consider light cone effects in this paper, although of course a real observation would include such effects \citep{Datta_2014,La_Plante_2014}. We fix our boxes to have $200 \times 200 \times 200$ voxels, with each side corresponding to $300\,\textrm{Mpc}$. In constructing our dataset, both cosmological and astrophysical parameters are set at their default values in \texttt{21cmFAST} (see, e.g., the fiducial values used in \citealt{Park2019}). In future work, it will be important to consider other parameter values; for this paper, however, we follow the precedent of \citet{li2019} and \citet{Makinen2020} and keep parameters fixed. Again, our goal is to provide a proof-of-concept study to establish that it is indeed possible to recover the morphology of ionized regions in wedge-filtered images. Our study is therefore highly complementary to that of \citet{Bianco}, who have trained a network for ionized bubble identification that is robust to a wide range of parameter values and have performed an extensive study of instrumental noise levels, but do not include the effects of foregrounds.

%
With the aforementioned \texttt{21cmFAST} settings, we generated a total of $20$ random realizations, each with different random seeds for the initial density field of the simulations. By producing simulation boxes at various redshifts between $z = 8.5$ and $z = 8.0$, we obtain a total of $57$ different boxes. The number of \texttt{21cmFAST} simulations
was limited to 57 due to the computational resources required
to train a 3D neural network on large amounts of data. While
these simulations are computationally cheap to realize, they
are not cheap for the network to train on. One of the random realizations is evolved down to redshift $z = 7.85$ so that its $z = 7.85$ realization is used only in validation. Other-redshift realizations of this seed are used in training. The motivation here was to test domain transfer across neutral fractions, i.e., to see to what extent a neural network trained in the range $z = 8.5$ to $8.0$ would work on a box at $z = 7.85$. The range of redshifts was held to this narrow
range due to the limited number of training boxes. Widening the
range would either mean generating more boxes, or sampling
boxes more sparsely across redshifts, the former of which is too computationally expensive and the latter provides too few
examples from each redshift for adequate network performance.

With pristine $21$-cm brightness temperature boxes on hand, we must corrupt the simulation data to reflect real-world instrumental and data analysis effects. In this paper, we consider three classes of data:
\begin{enumerate}
\item \textbf{Noiseless data.} Each $21$-cm brightness temperature cube is first Fourier transformed, and then all Fourier modes outside of the ``EoR window" are zeroed out. The result is inverse Fourier transformed to give a final box in configuration space. This represents what a perfect, noiseless instrument might see once foreground-contaminated wedge modes are excised.
\item \textbf{Noisy data.} In two parallel datasets, we included instrumental effects. As our fiducial instruments we consider HERA and the Square Kilometre Array (SKA; \citealt{Koopmans2015}). The motivation here is that HERA represents a current-generation instrument that is not necessarily optimized for imaging, whereas the SKA is a next-generation instrument that is better suited for imaging. For each of these interferometers, we take into account its Fourier-space sampling (i.e., the $uv$ distribution in radio astronomy parlance) to convolve the original $21$-cm brightness temperature boxes with an appropriate---and non-trivial---point spread function. The $uv$ distribution is determined by the antenna layout in the interferometer array. For HERA we assume its full 350-dish configuration, with 320 dishes in an ``split hexagon" layout and 30 outrigger dishes (see \citealt{DillonParsons2016,DeBoer:2017} for details). For the SKA we assume the fiducial design outlined in the ``SKA Admin - SKA TEL SKO DD 001 1 Baseline Design 1" memo\footnote{\url{https://www.skatelescope.org/ska-tel-sko-dd-001-1_baselinedesign1/}}.

We also use a modified version of \texttt{21cmSense}\footnote{\url{https://github.com/jpober/21cmSense}} \citep{Pober2013,Pober2014a} to add Gaussian random noise (according to the radiometer equation) to the sampled Fourier modes, thereby producing instrumental noise that has the proper pixel-to-pixel correlations in configuration space. After adding instrumental noise we perform the wedge excision as with the noiseless data (mimicking the sequence that would take place with real observations). A total integration time of $1080\,\textrm{hrs}$ is assumed. The preprocessing pipelines for the noiseless and noise-inclusive trainings are shown in Figure \ref{fig:preprocess_flowchart}, where the numbers in parentheses represent the size of a tensor at each step. One sees from the numbers that in many cases, it was necessary to tile the \texttt{21cmFAST} boxes in order to match the fact that HERA and the SKA have wide fields of view (relative to the angle subtended by a $\sim 300\,\textrm{Mpc}$ simulation box at $z \sim 8$). To eliminate---or at least mitigate---possible artifacts from the resulting periodicity, we extract a small box equal in dimension to the original boxes to feed into our neural networks, \emph{after} applying instrumental and noise effects.
\item \textbf{Null tests.} Finally, we consider a set of ``Gaussianized" boxes in order to test our hypothesis that it is non-Gaussian correlations between Fourier modes that enable the reconstruction of modes within the foreground wedge. If our guess is correct, an accurate reconstruction of the original images should fail. We Gaussianize our boxes in two ways. One method is to take the Fourier transform of each \texttt{21cmFAST} brightness temperature map and replace the phase of each Fourier coefficient with a phase drawn from a uniform distribution between $0$ and $2 \pi$ while preserving its amplitude. The second method is to generate a new Gaussian realization of a map given the power spectrum of the original (non-Gaussian) map, followed by an assignment of pixels from the new map to the old map by their ranking in brightness. In other words, the value of the dimmest pixel in the Gaussian realization replaces the value of the dimmest pixel in the original map, the second dimmest pixel replaces the second dimmest pixel in the original, and so on. In this way, a histogram of pixels in our map is Gaussian but we preserve the morphology of having low brightness temperature ``bubbles'' and higher brightness temperatures elsewhere. In both cases the network is trained on the ``Gaussianized" boxes before attempting to reconstruct them.
\end{enumerate}
%
%
%
%
All trainings conducted in this study used a collection of 57 brightness temperature boxes (pre-processed to include instrumental effects in the manner that we have just described). Of these, 51 were used for training our neural networks and 6 were used for validation. The distribution of the redshifts and neutral fractions of the training and validation sets are shown in Figure~\ref{fig:file-dist}. The inputs to our neural networks are wedge-filtered brightness temperature maps which have been normalized between 0 and 1. During training the outputs are compared to ground truth binarized maps where all non-zero voxels in $21$-cm brightness temperature are set to one (i.e. any voxel which is not fully ionized is considered fully neutral by the network). This binarization is performed to simplify the task into a two-class image segmentation problem where we are simply interested in knowing whether a part of our Universe is neutral or not.

\subsection{U-Net Architecture}

\begin{figure*}
    \centering
	\includegraphics[trim= 0 80 0 0, clip, width=7 in]{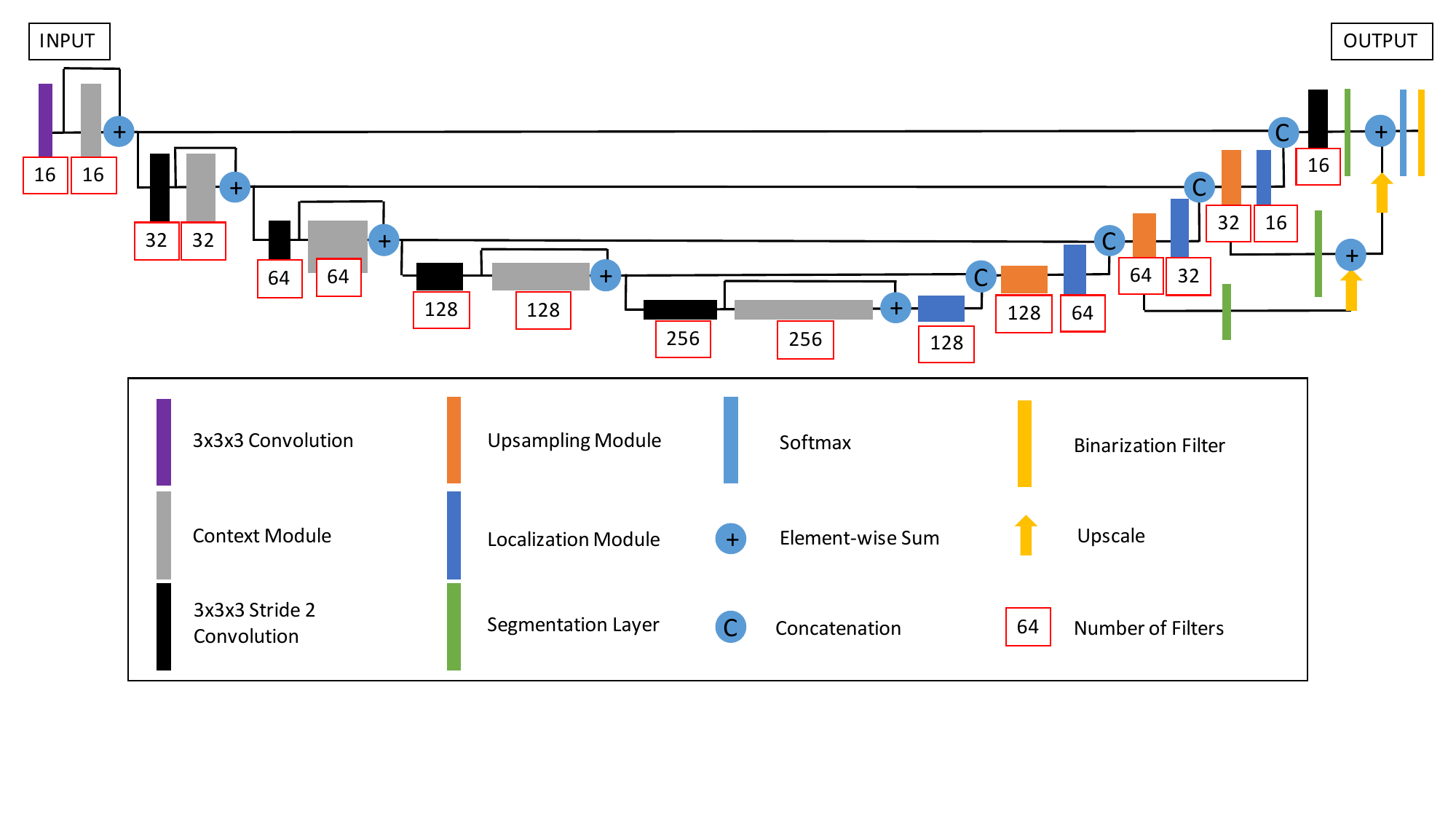}
	\caption{A block diagram of the U-Net used in this work.}
	\label{fig:isensee}
\end{figure*}

In its simplest form, our problem is one of image segmentation. 
We have an image wherein some regions are ionized and the rest is neutral, but the boundaries between these regions are not obvious after passing through the wedge filter.
A desirable wedge-removing network is able to label each pixel within the wedge-affected map as neutral or ionized, which is an image segmentation task. Given this, we select a U-net architecture for our neural network given the U-net's demonstrated success in image segmentation tasks \citep{Ronneberger2015,Isensee2019}. Our U-Net draws heavily from the architecture presented in \citet{Isensee2018}. In what follows we closely mirror the presentation in that paper, while also highlighting modifications made to the network for this work. 

A schematic of our neural network in shown in Figure~\ref{fig:isensee}. The network is configured to process large 3D input blocks
of $128\times 128\times 128$ voxels. These inputs are wedge-affected images normalized (not binarized) to range from 0 to 1. The basic U-Net architecture intrinsically recombines different scales throughout the entire network,
allowing it to make effective use of the entire input volume. The general U-Net architecture consists of a contextualization pathway (left branch) which encodes increasingly abstract representations of the input as one progresses deeper into the network, followed by a localization pathway (right branch) which recombines the abstract representations with shallower
features in order to precisely localize the structures of interest. The vertical depth in the U-Net is referred to as the level, with deeper
levels having lower spatial resolution and more channels than shallower levels.

The activations in the context pathway are computed by a pre-activation residual block containing two $3\times 3\times 3$ convolutional layers with a spatial 
dropout layer in between. These are referred to as context modules, and are shown in grey in Figure \ref{fig:isensee}. Unlike in \citet{Isensee2018}, we employ spatial dropout instead of normal dropout to improve regularity and training 
stability \citep{tompson2014efficient}. Spatial dropout shuts off entire feature maps rather than individual neurons, meaning that adjacent pixels
in the post-dropout feature map are either all 0 or all active. This helps reduce overfitting while still allowing the network to contextualize
spatial information. Each pre-activation residual block is connected by a $3\times 3\times 3$ convolutional layer with stride 2 (shown in black) to reduce the resolution of the
feature maps. Stride is used instead of pooling since we found it to have superior performance.

The localization path consists of successive localization and upsampling modules alongside a parallel path which facilitates deep supervision.
Each upsampling module (shown in orange) is a $2\times 2\times 2$ upscaling operation which tessellates the feature voxels twice in each spatial dimension, followed by a 
$3\times 3\times 3$ convolution that halves the number of feature maps. Upscaling was chosen over the more popular transposed convolution since it was found by \citet{Isensee2018} to 
prevent checkerboard artifacts in the network output. Our own tests corroborate this assessment. The upsampled features are then 
concatenated with the features from the corresponding level of the contextualization pathway. A localization module then combines these features
together. Each localization module (shown in blue) consists of a $3\times 3\times 3$ convolution followed by a $1\times 1 \times 1$ convolution that halves the number of feature maps.

Deep supervision is employed in the localization pathway by taking a segmentation layer (shown in green) after each localization module, upscaling the segmentation map
by a factor of two, and then adding it elementwise to the next segmentation map. Thus the final output of the network integrates information from segmentation
maps made at all levels of the network. All feature map computing convolutions use leaky ReLU activation functions with a negative slope of $10^{-2}$.
Instance normalization is used on all contextualization modules instead of batch normalization since the stochasticity induced by small batch sizes can destabilize batch normalization
\citep{Isensee2018,ulyanov2016instance}. Skip connections connect layers of equal depth across the network via concatenation along the channel axis, as per the original U-Net design presented in \citet{Ronneberger2015}.


The final layer of the network is a so-called ``binarization filter", which maps each voxel in the output to zero or one depending on some threshold. It is not used during training in order to incentivize the network to produce near-binary outputs. When predictions are generated for post-training testing, the binarization filter is used with a threshold of $0.9$. Some level of arbitrariness exists in the determination of the cutoff used in binarizing the prediction and ground truth boxes. We selected $0.9$ to provide conservative ionized regions. However, we found that variations in threshold between $0.5$ and $0.9$ did not significantly change the prediction maps.

Our network is trained on 51 input images with a batch size of 3. We refer to an iteration over 26 batches as an epoch and train for a total of 100 epochs. Training is done using an \texttt{Adam} optimizer with an initial learning rate of ${lr}_\textrm{init}=5\cdot10^{-4}$, and an exponentially decaying learning schedule (${lr}_\textrm{init}\cdot0.985^{E}$, where $E$ is the number of epochs elapsed). The network is trained using a differentiable approximation of the binary dice coefficient function, defined as
\begin{equation}
    D_{loss} \equiv 1 - \frac{2\times|X\cap Y| + \alpha}{|X| + |Y| + \alpha},
\end{equation}
where $X$ and $Y$ represent the truth and prediction matrices, respectively, and $\alpha$ represents a small number used to avoid divide-by-zero errors (in our implementation, $\alpha=1$). The binary dice coefficient is a proxy for the intersection-over-union statistic commonly used to evaluate the performance of image segmentation algorithms. Although the binary dice loss is more computationally expensive than the widespread binary cross-entropy loss function, we selected it anyway since it is more optimized for image segmentation tasks \citep{milletari2016vnet}.

\section{Results}
\label{sec:results}

\subsection{Training}
\label{sec:training}

\begin{figure}
    \centering
	\includegraphics[width=0.45 \textwidth]{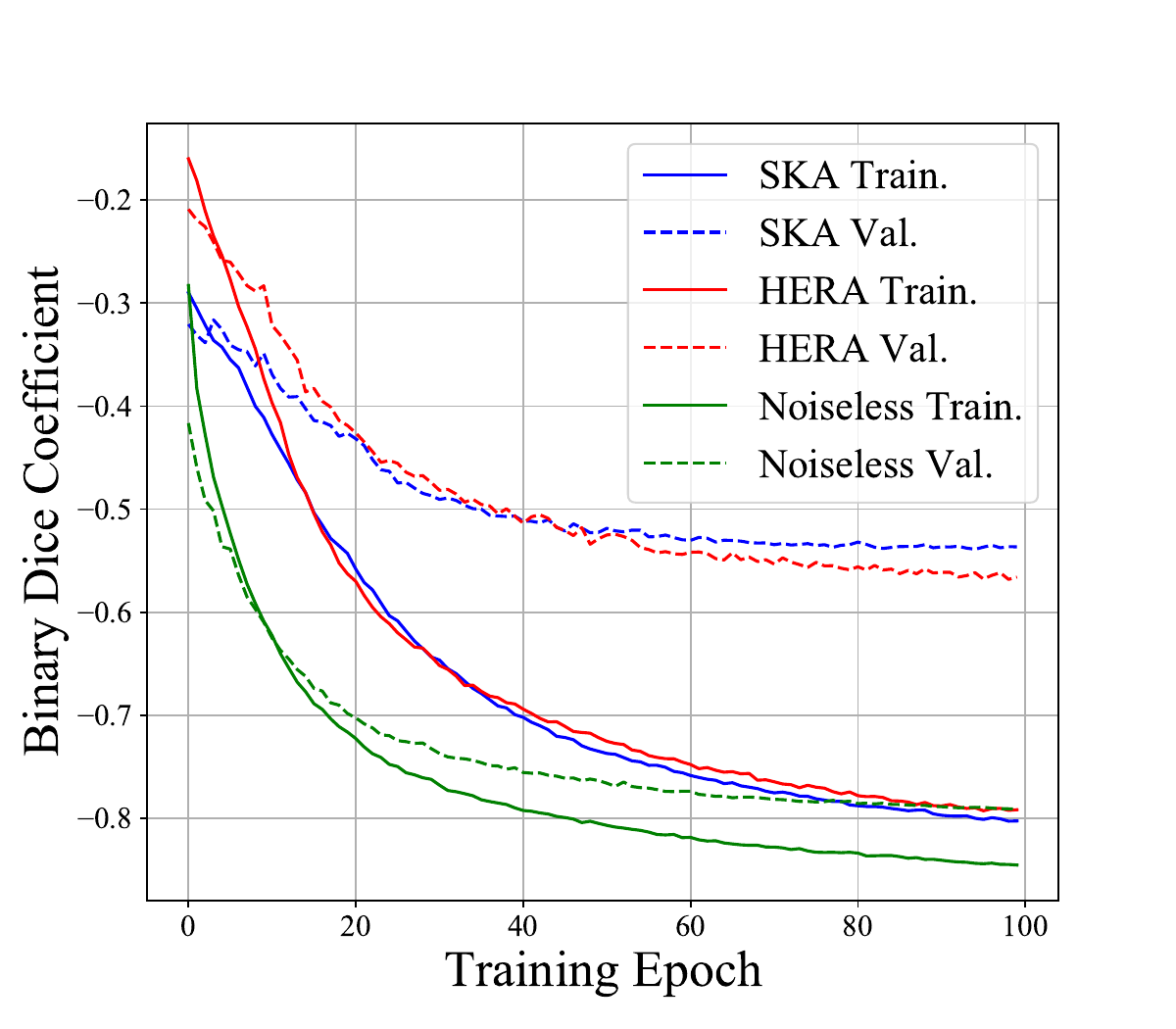}
	\caption{The binary dice loss computed at each epoch of training for each of the three models. Solid lines are used for training loss, while the dashed lines are used for validation loss. The curves are indicative of learning without significant over-fitting.}
	\label{fig:loss}
\end{figure}

The binary dice coefficients calculated for training and validation data at the end of every epoch of training are shown in Figure \ref{fig:loss}. In none of the three models is a point reached in training where the training loss continues to decrease while the validation loss increases. This suggests that our network is not over-fitting. Furthermore, by 100 epochs all validation loss curves have entered a domain of near-flatness, indicating that the network has learned all that it can from the data set. However, in all three models a large divide separates the validation loss from the training loss, possibly indicating that our learning may benefit from a training set of larger size or variation \citep{ANZANELLO2011}.

\subsection{Network Predictions}

\begin{figure*}
    \centering
    \includegraphics[width=6 in]{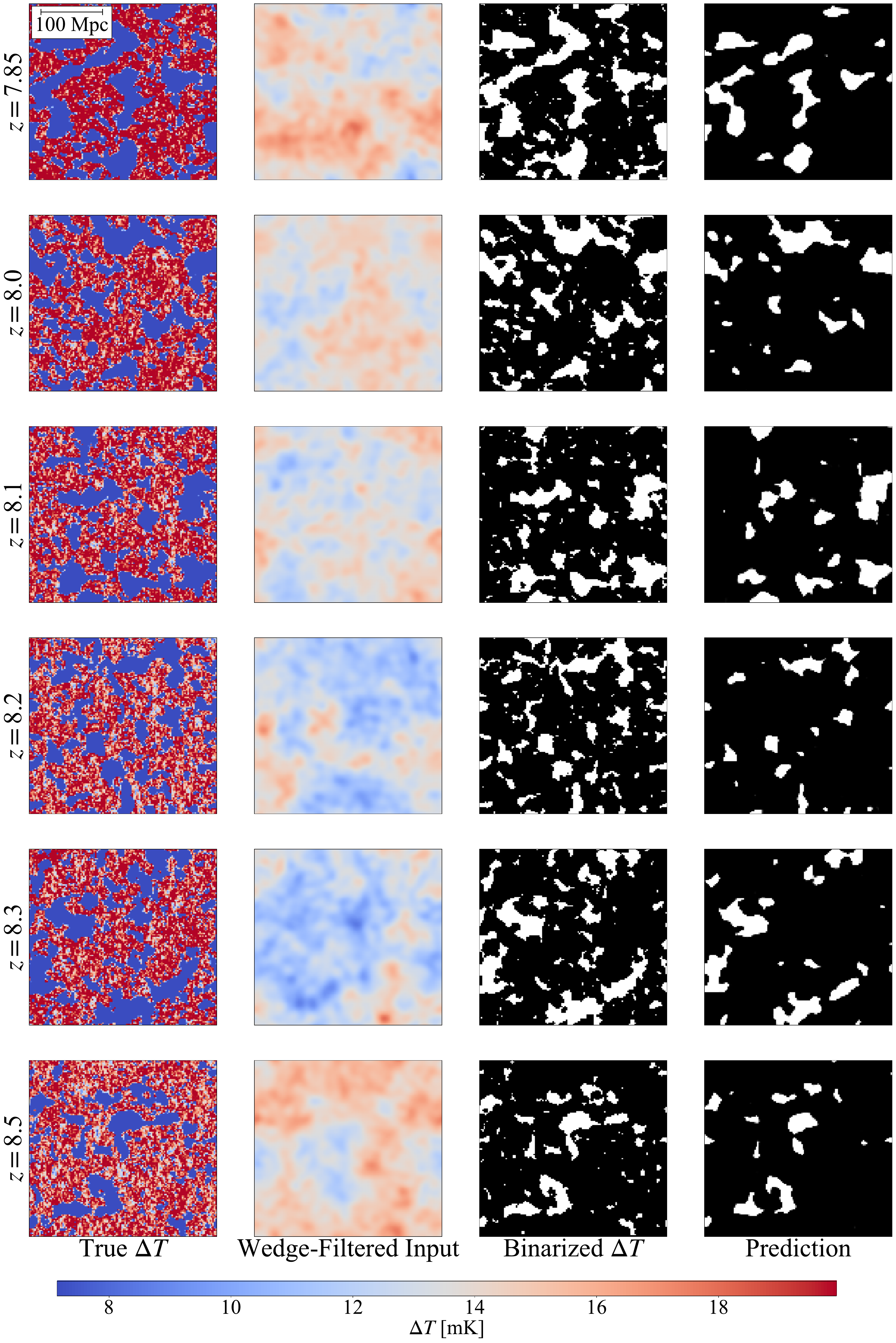}
    \caption{Sample network predictions on noiseless data suite. The first column shows a transverse cross-section of the true brightness temperature field, while the second column shows the same field after excising Fourier modes lying within the foreground wedge. The third column is a binarized version of the first column and serves as the ground truth for our neural network. The fourth column shows the predicted ionization maps from our network. Visually, it is clear that our network is able to recover ionized bubbles from wedge-filtered $21$-cm maps.}
    \label{fig:pred-noiseless}
\end{figure*}

\begin{figure*}
    \centering
    \includegraphics[width=6 in]{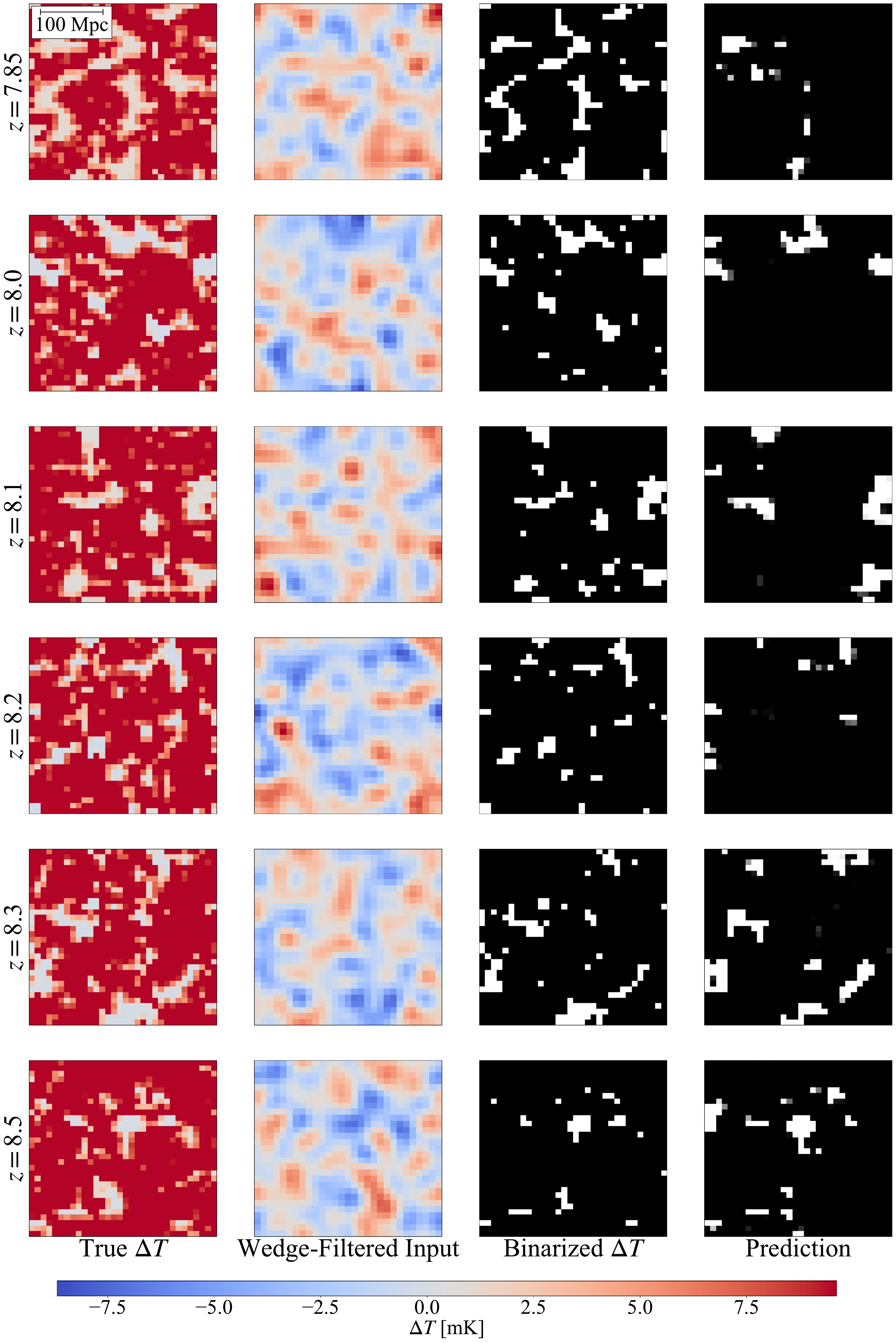}
    \caption{Same as Figure~\ref{fig:pred-noiseless} except with the HERA data suite.}
    \label{fig:pred-HERA}
\end{figure*}

\begin{figure*}
    \centering
    \includegraphics[width=6 in]{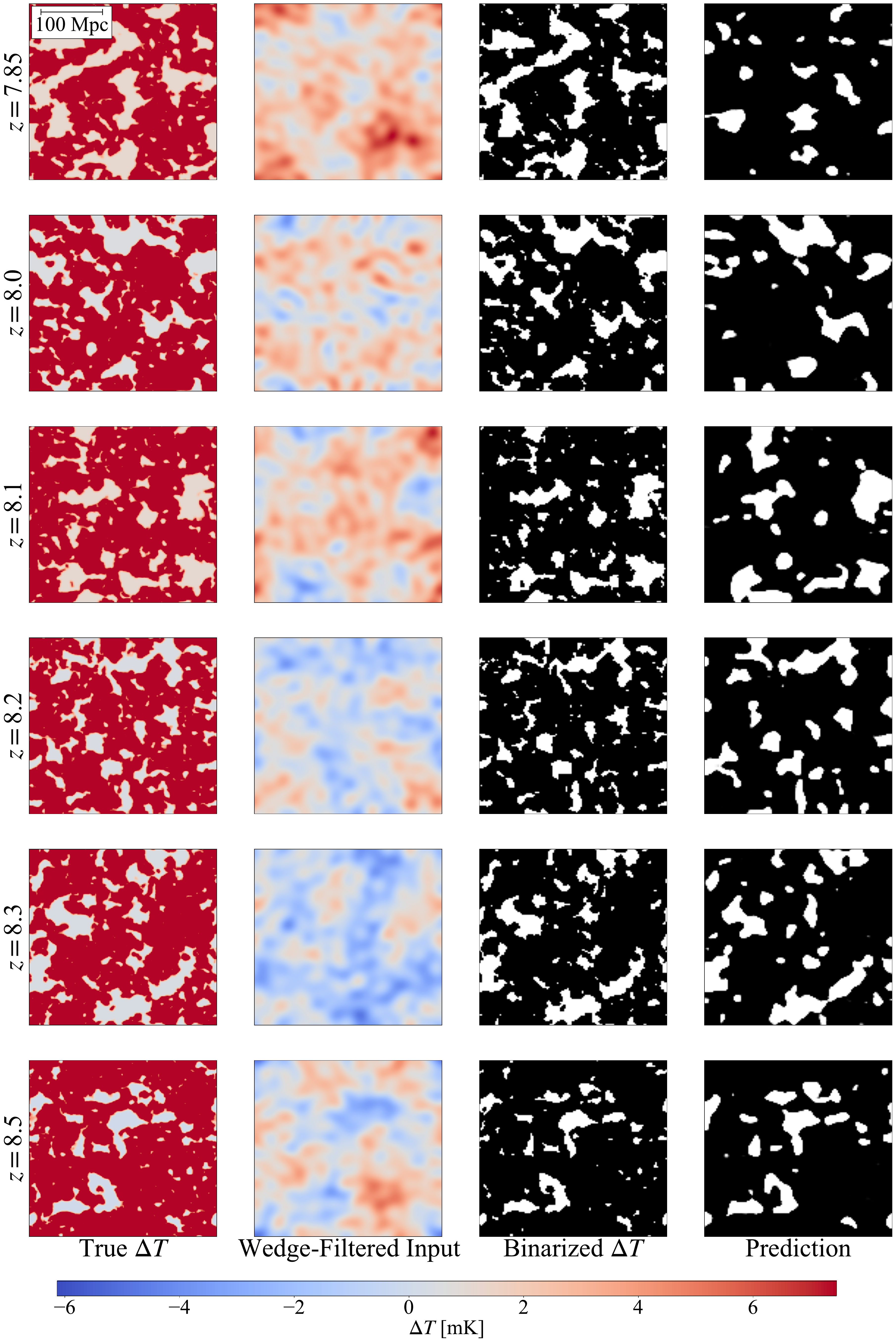}
    \caption{Same as Figure~\ref{fig:pred-noiseless} except with the SKA data suite.}
    \label{fig:pred-SKA}
\end{figure*}

\begin{figure*}
    \centering
	\includegraphics[trim=100 0 100 0, clip, width=\textwidth]{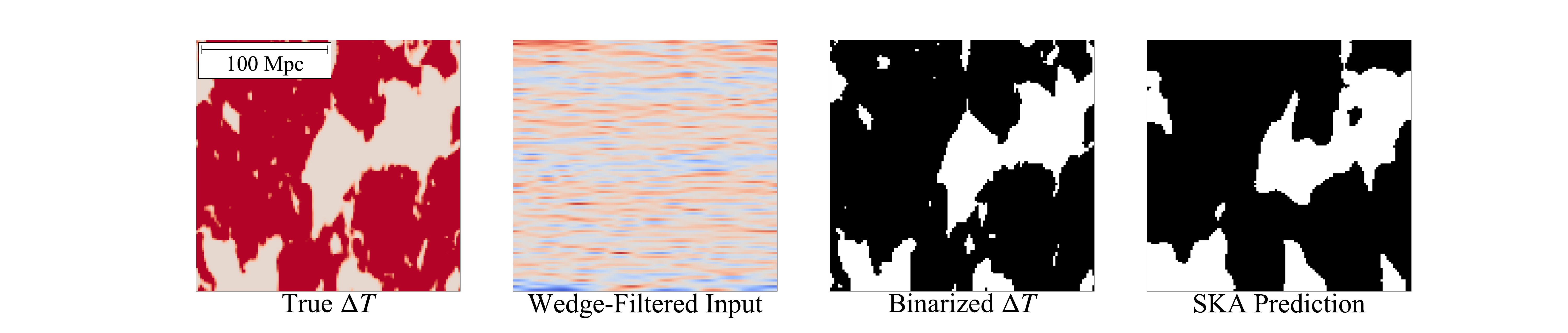}
	\caption{Same as the top row of Figure~\ref{fig:pred-SKA}, except with image slices containing one transverse axis and the line-of-sight axis. One sees that while the reconstruction of the $z=7.85$ bubbles appears to be poor in Figure~\ref{fig:pred-SKA} when we just consider transverse slices, many line-of-sight structures are predicted correctly, suggesting a reasonable overall 3D reconstruction.}
	\label{fig:los-SKA}
\end{figure*}

Figures \ref{fig:pred-noiseless}, \ref{fig:pred-HERA}, and \ref{fig:pred-SKA} display sample predictions from each validation box in each test. Each figure is arranged into four columns and six rows. The first column in each figure shows a cross-section of the original 21-cm brightness temperature map. In Figures \ref{fig:pred-HERA} and \ref{fig:pred-SKA} this temperature map is sampled by HERA and the SKA's Fourier footprints, respectively. Appropriately correlated noise is added, in accordance with the procedure outlined in Section~\ref{sec:ForwardSim}. The second column of each figure shows a cross-section of the wedge-filtered input to our network. The third column shows the 21-cm temperature field after being passed through a binarization filter; it is this column that represents the ground truth that our algorithm is trying to reproduce. The final column shows the prediction made by the network. Each row shows a sample set from one of the redshifts included in the validation data set. In all figures the arbitrary decision is made to show cross-sections which are perpendicular to the line of sight direction. As we know from Figure~\ref{fig:wedge_demo}, slices along the line of sight direction look substantially different and contain unique information. We remind the reader that our network takes in 3D data cubes and outputs 3D data cubes, and thus all of this information is used in the prediction.

Figure \ref{fig:pred-noiseless} displays sample predictions from the noiseless model. Comparing the third and fourth columns, it is clear that the network is capable of reproducing the sizes, shapes, and locations of the largest bubbles in each image. However, it is also evident that many structures present in the ground truth do not appear in the prediction, especially small structures. While the network misses many structures, it does not tend to create structures which are not present in the ground truth. This observation will be expanded upon as we discuss the prediction statistics. The performance of the network does not appear to be significantly better or worse at any redshift.

Figure \ref{fig:pred-HERA} displays sample predictions from the HERA model. Despite HERA's low resolution, the network still captures the locations of the major ionized regions, and in all redshifts except for $z=7.85$ it is able to reproduce the size and shape of the largest few bubbles. This opens the door to the limited imaging work which can be done using HERA, which was intended as a primarily statistical measurement experiment.

Figure \ref{fig:pred-SKA} displays sample predictions from the SKA model. Since the SKA is an instrument more optimized for imaging, its predictions are near in fidelity to those in the noiseless case. As with the previous two cases, the network neglects the smallest bubbles in favour of the largest. Similarly to the HERA case, the SKA model performs more poorly on redshift $z=7.85$ than on other redshifts. However, we note that the seemingly poor performance here is in fact a visual artifact of our plotting a transverse slice of the data cubes. Figure~\ref{fig:los-SKA} shows slices with one transverse and one light-of-sight axis. It is visually apparent that many of the ionized bubble structures are recovered along the line of sight. This suggests that even if the network does not perform quite as well when validated on boxes from redshifts that were not used in training (recall Figure~\ref{fig:file-dist}), there is still some degree of success when considering the predictions in a three-dimensional volume.

\subsection{Prediction Statistics}

\begin{table}
\centering
\begin{tabular}{|l|l|l|}
\hline
\textbf{Prediction} & \textbf{Ground Truth} & \textbf{Class} \\ \hline
Ionized & Ionized & True Positive\\ \hline
Ionized & Neutral & False Positive\\ \hline
Neutral & Ionized & False Negative\\ \hline
Neutral & Neutral & True Negative\\ \hline
\end{tabular}
\caption{The logic scheme used to determine the class of a prediction voxel. These class labels are then used to calculate the accuracy, precision, recall, and intersection-over-union metrics.}
\label{tab:class-defs}
\end{table}

\begin{table}
\centering
\begin{tabular}{|l|l|l|l|l|}
\multicolumn{5}{c}{\textbf{Noiseless}}                           \\ \hline
\textbf{Neutral Fraction} & \textbf{Accuracy} & \textbf{Precision} & \textbf{Recall} & \textbf{IoU} \\ \hline
0.474 & \textit{0.823} & 0.987 & \underline{0.508} & \underline{0.504} \\ \hline
0.522 & 0.843 & 0.987 & 0.482 & 0.479 \\ \hline
0.544 & 0.861 & \textit{0.977} & 0.507 & 0.501 \\ \hline
0.593 & 0.86 & 0.992 & \textit{0.364} & \textit{0.363} \\ \hline
0.615 & 0.879 & \underline{0.995} & 0.411 & 0.41 \\ \hline
0.656 & \underline{0.896} & 0.986 & 0.388 & 0.386 \\ \hline
\multicolumn{5}{c}{\textbf{HERA}} \\ \hline
\textbf{Neutral Fraction} & \textbf{Accuracy} & \textbf{Precision} & \textbf{Recall} & \textbf{IoU} \\ \hline
0.474 & 0.861 & \underline{0.542} & \textit{0.479} & 0.341 \\ \hline
0.522 & 0.907 & 0.527 & 0.623 & 0.4 \\ \hline
0.544 & 0.901 & 0.365 & 0.581 & \textit{0.289} \\ \hline
0.593 & 0.948 & 0.488 & 0.658 & 0.389 \\ \hline
0.615 & \underline{0.975} & 0.485 & 0.73 & \underline{0.411} \\ \hline
0.656 & \textit{0.974} & \textit{0.329} & \underline{0.812} & 0.306 \\ \hline
\multicolumn{5}{c}{\textbf{SKA}} \\ \hline
\textbf{Neutral Fraction} & \textbf{Accuracy} & \textbf{Precision} & \textbf{Recall}   & \textbf{IoU} \\ \hline
0.474 & \textit{0.792} & \underline{0.771} & \textit{0.507} & 0.441 \\ \hline
0.522 & 0.837 & 0.701 & \underline{0.623} &  0.492 \\ \hline
0.544 & 0.817 & \textit{0.582} & 0.609 & \textit{0.424} \\ \hline
0.593 & 0.881 & 0.701 & 0.607 & 0.482 \\ \hline
0.615 & 0.925 & 0.755 & 0.587 & \underline{0.493} \\ \hline
0.656 & \underline{0.928} & 0.649 & 0.609 & 0.458
\end{tabular}
\caption{The tabulated statistics for the predictions made by the network on each validation data suite. The highest score in each column is underlined, while the lowest is italicized.}
\label{tab:all-stats}
\end{table}

The network's performance in each test is evaluated on the similarity of the validation prediction data to their corresponding binarized ground truth data. This is judged for the first three models using the accuracy, precision, recall, and intersection-over-union (IoU) statistics. The first three metrics are calculated by classifying each voxel of a prediction box into one of four classes: true positives, true negatives, false positives, and false negatives. The logic scheme used for class assignment is shown in Table \ref{tab:class-defs}. These are then distilled into scores by taking the number of voxels in each class for a given box and dividing by the total number of voxels in the box. For example, if a box with a resolution of $128\times128\times128$ has $1500000$ ``false positive" voxels, then its ``false positive" score is $0.715$. The sum of all four scores for any box is $1$. In what follows, we will denote the true positive score as TP, false positive as FP, false negative as FN, and true negative as TN.

Accuracy is a measure of the overall prediction fidelity. It is defined as 
\begin{equation}
    \mathrm{Accuracy}=\frac{\mathrm{TP}+\mathrm{TN}}{\mathrm{TP}+\mathrm{FP}+\mathrm{FN}+\mathrm{TN}}.
\end{equation}
Since accuracy accounts for the populations of all four classes, it is easily inflated in situations where one class is overwhelmingly present. For example, if a validation box is $99.9\%$ neutral, and the network improperly identifies the $0.1\%$ region which is ionized, then the accuracy of the prediction will be $99.8\%$ despite the network not properly labelling a single ionized voxel. Therefore, other metrics are necessary in order to capture full texture of a network's classification biases.

Precision is a measure of how many voxels labelled as ionized by the network are truly ionized. It is defined as
\begin{equation}
    \mathrm{Precision}=\frac{\mathrm{TP}}{\mathrm{TP}+\mathrm{FP}}.
\end{equation}
This is useful in situations where the ``cost" of a false positive is high. In our study, we want to make sure that our network is predicting ionized regions that actually exist, with an eye towards future studies where ionized regions from $21$-cm maps can be used to direct searches for high-redshift galaxies.

Recall is the share of truly ionized voxels which are labelled as ionized in the prediction. It is defined as
\begin{equation}
    \mathrm{Recall}=\frac{\mathrm{TP}}{\mathrm{TP}+\mathrm{FN}}
\end{equation}
A highly conservative network will have low recall, since it only labels regions which it is highly confident in as positive. Such a network may properly locate the rough location of ionized bubbles, but may not accurately portray their size or morphology by being too conservative about pixels on the edge of the bubbles.

IoU is a measure of the overlap between a prediction and its ground truth, defined as the algebraic intersection between two boxes divided by their union. It is commonly used to evaluate the predictions of image segmentation neural networks \citep{rezatofighi2019generalized}, and is included in this study for ease of comparison with similar networks. IoU is calculated via
\begin{equation}
    \mathrm{IoU}=\frac{|P\cap T|}{|P\cup T|},
\end{equation}
where $P$ is the binarized prediction $P$ and $T$ is the binarized ground truth. Both are 3-dimensional boolean arrays.

These statistics are tabulated for each box in the validation set in Table \ref{tab:all-stats} which contains the results for the noiseless, HERA, and SKA models. The validation boxes are identified by their neutral fractions. What follows is a discussion of the statistics of each model's predictions, and what they may imply about the tendencies of each model.

While none of the statistics for the noiseless test are strongly correlated with the neutral fraction of the box, it is perhaps notable that the network performed best in the recall and IoU statistics on the three boxes with the lowest neutral fractions. This is probably not the result of a bias in training, since the training set is more heavily biased towards large neutral fractions (see Figure \ref{fig:file-dist}). Notably, the box with the highest IoU also has the lowest accuracy, indicating that the network is likely to mark ionized voxels as neutral, but unlikely to mark neutral voxels as ionized. 

The HERA statistics are oriented with the highest recall and IoU statistics lying on the higher end of the neutral fraction spectrum and the highest precision on the lowest neutral fraction box. It is notable that the HERA boxes have a much lower resolution than the noiseless or SKA boxes, so less small-scale detail exists to be mined in the first place. This could have the effect of suppressing recall on low-neutral fraction boxes, where the ionized bubbles tend to be smaller.

The SKA statistics are comparable in their distribution to the HERA statistics, save for recall, which does not vary as greatly among neutral fractions as in the HERA case. The precision scores are higher than any on the HERA prediction, but they fall short of the noiseless predictions. Since precision is a measure of the number of true positives out of all pixels labelled positive, this is likely a matter of image resolution. The HERA model does not have the liberty to set a low confidence threshold for labelling each pixel since it has relatively few to work with. Meanwhile, the SKA and noiseless models have high resolution images, and can afford to scrutinize each pixel.

\begin{figure}
    \centering
    \includegraphics[trim= 0 50 0 0, clip, width=\columnwidth]{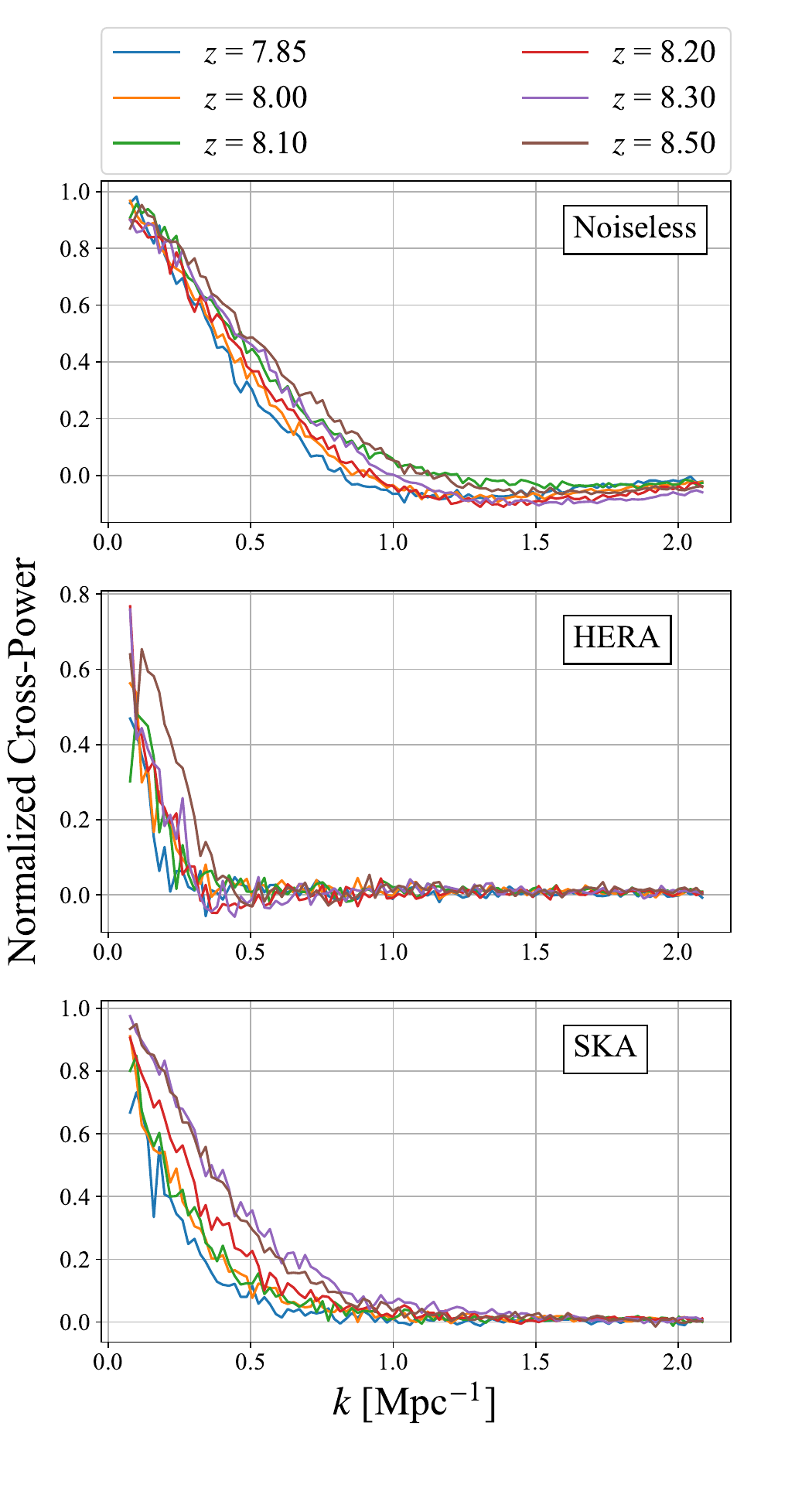}
    \caption{The normalized cross-power between each prediction and its associated binary mask in the validation sets for the noiseless, HERA, and SKA models. In all cases, normalized cross-power is highest at low $k$, indicating that the network is better at recovering large-scale structures than small-scale structures. This fidelity drop-off with spectral scale is referred to in machine learning literature as ``spectral bias" \citep{rahaman2019}.}
    \label{fig:crosscor}
\end{figure}

\begin{figure}
    \centering
    \includegraphics[trim= 0 50 0 0, clip, width=\columnwidth]{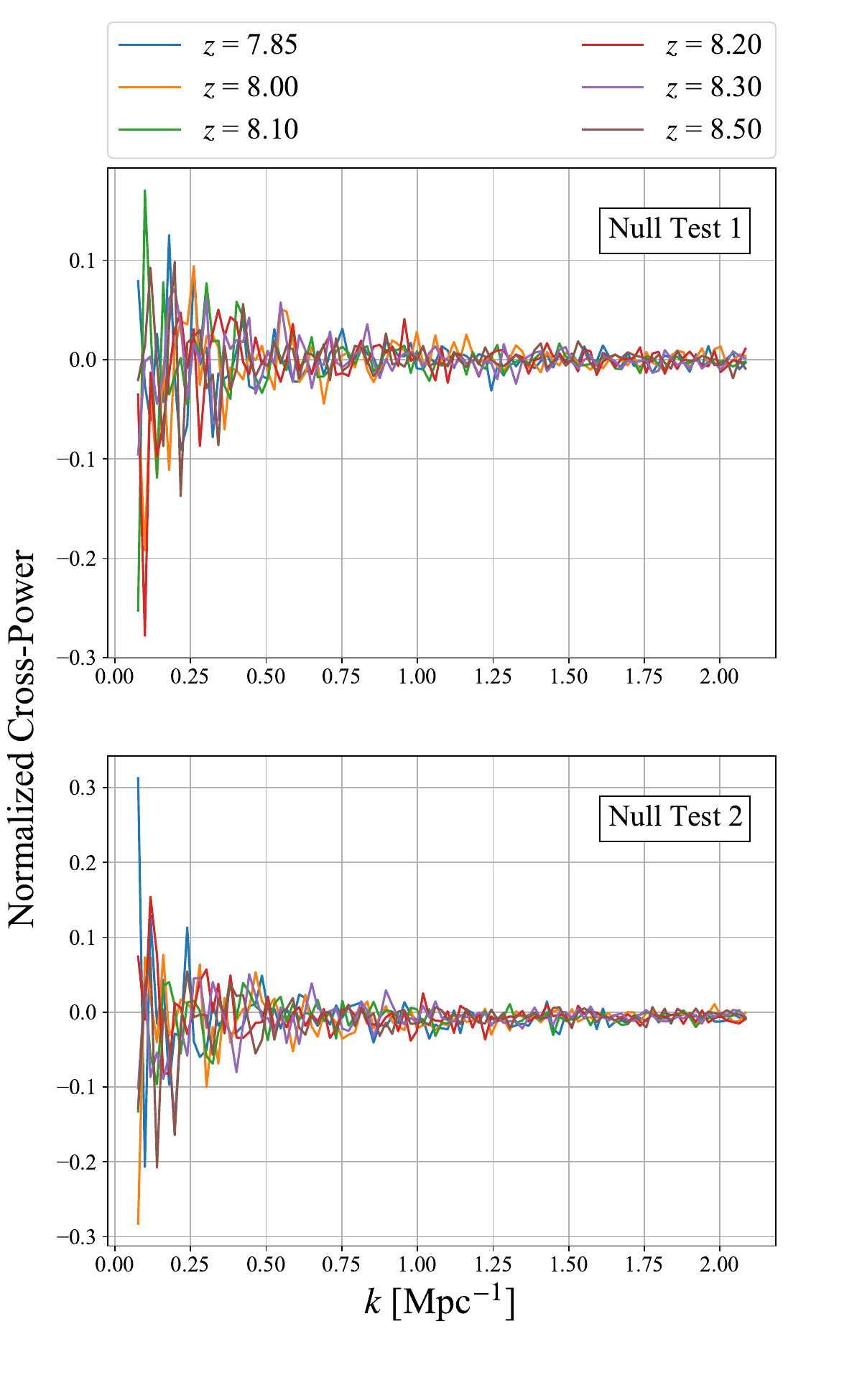}
    \caption{The normalized cross-power between each prediction and its associated binary mask in the validation sets for both null test models. In both cases, the normalized cross-power is near zero across $k$ modes.}
    \label{fig:nullcor}
\end{figure}

\subsection{Cross-Power Spectra}

\begin{figure*}
    \centering
    \includegraphics[trim = 80 0 0 80, width=6 in]{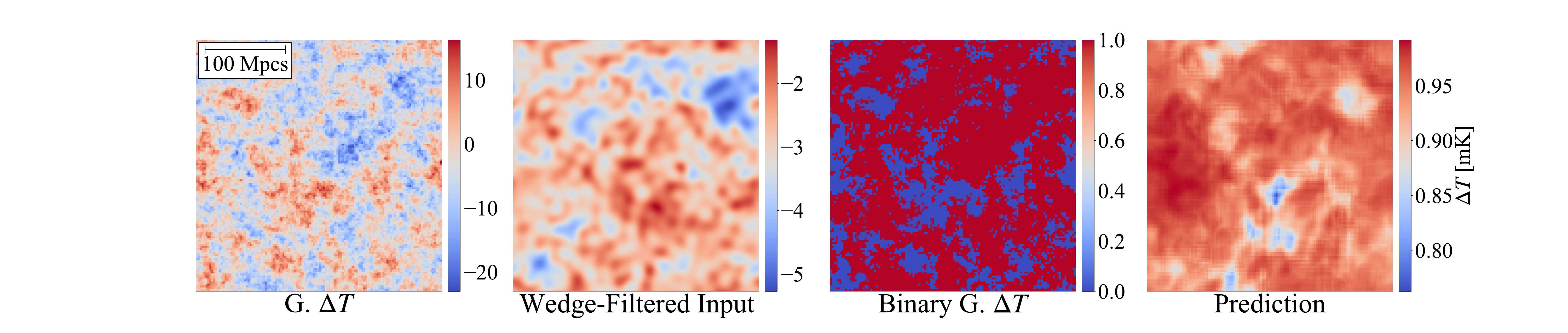}
    \caption{Sample network prediction on a validation box slice from $z=7.85$ in the first null test data suite. The network fails to reproduce any meaningful structures present in the ground truth image. As with Figure~\ref{fig:pred-noiseless}, the leftmost image shows the $21$-cm brightness temperature field, except here the field has been Gaussianized. The second image is the same as the first image after a wedge filter. The third image is a binarized version of the first, and the rightmost image is the (failed) attempt at predicting the third image. This confirms our intuition that the key to reconstructing Fourier modes lost to the wedge is the non-Gaussianity of EoR maps.}
    \label{fig:pred-null}
\end{figure*}

\begin{figure*}
    \centering
    \includegraphics[trim = 80 0 0 80, width=6 in]{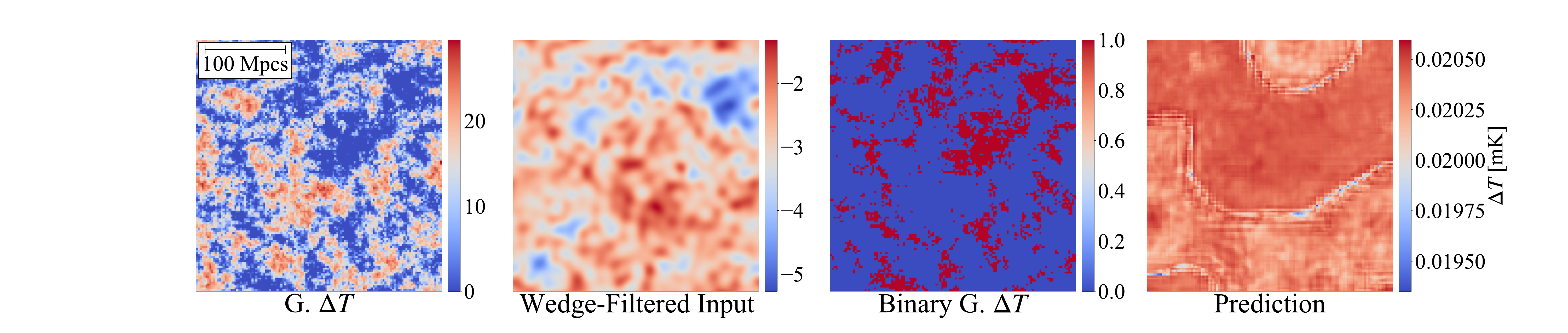}
    \caption{Same as Figure~\ref{fig:pred-null}, but for the second null test data suite.}
    \label{fig:pred-null2}
\end{figure*}

The normalized cross-power spectrum between the prediction and binarized mask was also calculated for each test. To define this cross-power, let us denote the prediction and binarized masks as $f$ and $g$, which are real-valued data sets. Let the Fourier transforms of $f$ and $g$ be $\mathcal{F}$ and $\mathcal{G}$, and the complex conjugates of these be $\mathcal{F}^*$ and $\mathcal{G}^*$. We define the normalized cross-power of $f$ and $g$ to be the power spectrum of

\begin{equation}
    \mathcal{N}=\frac{\mathcal{F}^*\mathcal{G}}{\sqrt{(\mathcal{F}^*\mathcal{F})(\mathcal{G}^*\mathcal{G})}},
\end{equation}

\noindent which is a complex-valued function of $k$. If $f$ and $g$ are identical, then the cross-spectrum is 1 at all $k$, and if they share nothing at all in common, then the cross-spectrum is 0 at all $k$. In this way, the cross-spectrum demonstrates the fidelity with which the network recovers different $k$-modes of an image.

Figure \ref{fig:crosscor} shows the normalized cross-power spectra for the noiseless, HERA, and SKA model predictions, while Figure \ref{fig:nullcor} shows the normalized cross-power spectra for both null tests. Common among the predictions in Figure \ref{fig:crosscor} is that the normalized cross-power drops off as a function of $k$. It does so most slowly for the noiseless suite, and most quickly for the HERA suite, suggesting a relationship between box resolution and prediction fidelity. The relationship between prediction fidelity and spatial frequency scale is a well-documented phenomenon in machine learning, referred to in the literature as spectral bias (for an in-depth discussion, see \citealt{rahaman2019}). In brief, spectral bias is the tendency for image reconstruction neural networks to perform better at low $k$-modes than at high $k$-modes. It is thought to arise from the granularity of details at high spectral scales, which makes them harder for a network to retrieve than large-scale ``generic" features. However, it is unclear to what extent spectral bias places a limit on the performance of the network. It is possible, for example, that our network is not optimally configured and still falls short of the fidelity limit imposed by the U-Net's spectral bias.

The demonstrated drop-off in fidelity at high $k$ illustrates that our algorithm is best suited for enabling image-associated science that relies on the identification of ionized bubbles. While it is not appropriate for improving measurements of the power spectrum or other Fourier space statistics, the network demonstrably excels at recovering the locations and sizes of ionized regions.

It is evident in Figure \ref{fig:crosscor} that the normalized cross-power at any given $k$ tends to increase with redshift, regardless of the noise type. This is probably an artifact of the training set, which leans strongly towards high-redshift boxes (see Figure \ref{fig:file-dist}). The effect may be further exacerbated on the $z=7.85$ line, since the networks were not trained on data from that redshift.

Meanwhile, the normalized cross-power spectra for both null test validation suites are very noisy and do not demonstrate any clear trend, besides being noisier at low $k$. This indicates that the network is completely unable to reconstruct the signal beneath the wedge when it is ``Gaussianized", supporting our hypothesis that the network is exploiting the non-Gaussian coupling between Fourier modes to reconstruct the EoR signal. This is confirmed by a visual inspection of the predicted images, shown in Figures \ref{fig:pred-null} and \ref{fig:pred-null2} for the first and second null tests described in Section~\ref{sec:ForwardSim}, respectively.

\section{Conclusions}


We have developed a machine learning-based method to identify ionized bubbles during the Epoch of Reionization. 
Our method considerably extends the work of \citet{li2019} and \citet{Makinen2020} and uses a U-Net-based deep learning algorithm to recover Fourier modes that are obscured by foregrounds.
The algorithm does not rely on any knowledge of the foregrounds themselves, and enables image reconstruction after all modes lying within the foreground wedge have been completely nulled out.
This is possible due to the significant non-Gaussian correlations between Fourier modes \citep{Shimabukuro_2016, Majumdar_2018, Watkinson_2018, Hutter_2019, Gorce_2019}.

Our main goal was to assess whether or not enough information exists in a wedge-filtered EoR image to reconstruct the original image within a reasonable margin of error. This paper demonstrates an affirmative answer to this question: the lost wedge modes can indeed be recovered from a wedge-filtered image by exploiting the non-Gaussian nature of the 21-cm EoR signal. We verify that the U-Net relies on phase correlations in the 21-cm signal by performing two null tests where the phases are decorrelated with one another. In both null tests, the U-Net fails to reconstruct any meaningful information (see Figures \ref{fig:pred-null} and \ref{fig:pred-null2} for sample predictions, and Figure \ref{fig:nullcor} for the Fourier space recovery fidelity in these null tests).

Additionally, we aimed to show that our methods remain viable when instrumental effects are accounted for, using HERA and the SKA as fiducial instruments. These instruments were selected since HERA is a current-generation instrument not necessarily optimized for imaging, while the SKA is a next-generation instrument more suitable for imaging. We found that the reconstruction fidelity in Fourier space drops off strongly as a function of $k$ (shown in Figure \ref{fig:crosscor}) and that better mode reconstruction will likely be necessary if one wishes to use our techniques for applications such as power spectrum estimation. However, in the image domain, the largest ionized regions in a wedge-filtered image can be reliably identified, even when the images include instrumental affects from HERA or the SKA (see Figures \ref{fig:pred-noiseless}, \ref{fig:pred-HERA}, and \ref{fig:pred-SKA} for sample predictions in the noiseless, HERA, and SKA cases). This demonstrates the capacity of even current-generation instruments like HERA to perform some limited imaging work, and paves the way for future EoR imaging studies.

In this paper, we have shown that filtering out foreground-contaminated modes within the wedge is \emph{not} a dealbreaker for imaging studies that seek to locate ionized bubbles during the EoR---the modes can be recovered to a sufficient extent using a neural network that in the image domain, the bubbles can be reliably identified. While our proof-of-concept study is an important first step, future work must considerably generalize our approach in order for it to be a practical tool. For instance, in this paper we kept astrophysical and cosmological parameters fixed, which does not accurately reflect our current state of knowledge in EoR studies. Progress has been recently made in this direction in a complementary study by \citet{Bianco} who have also tackled the problem of the EoR bubble identification over a wide range of parameter choices and instrumental noise scenarios, but not in the context of foreground filtering. Synthesizing these and other preliminary studies will allow 21-cm machine learning techniques to mature and take EoR imaging studies to the next level, unlocking the potential of 21-cm cosmology to even more dramatically alter our view of Cosmic Dawn than with just statistical studies alone.

\section*{Acknowledgments}

The authors are delighted to acknowledge helpful discussions with James Aguirre, Joelle Begin, Youssef Bestavros, Michele Bianco, Razvan Ciuca, Sambit Giri, Brad Greig, Nick Kern, Ilian Iliev, Paul La Plante, Garrelt Mellema, Andrei Mesinger, Damien Pinto, Jonathan Pober, Clovis Vinant-Tang, and Chris Williams. YC was funded by the Mitacs Globalink Research Internship Program. AL and SR are grateful for support from the Natural Sciences and Engineering Research Council of Canada (NSERC) through their Discovery Grants program as well as the Canadian Institute for Advanced Research (CIFAR) via the Azrieli Global Scholars program for AL and the Canada CIFAR AI Chair program for SR. Additionally, AL acknowledges support from the New Frontiers in Research Fund Exploration grant program, a NSERC Discovery Launch Supplement, the Sloan Research Fellowship, and the William Dawson Scholarship at McGill. Computations were made on the supercomputers Cedar (at Simon Fraser University) and B\'{e}luga (at \'{E}cole de technologie sup\'{e}rieure) managed by Compute Canada. The operation of this supercomputer is funded by the Canada Foundation for Innovation (CFI).

\section*{Data Availability}
The data underlying this article is available upon request. All 21-cm temperature anisotropy realizations can be re-generated from scratch using the publicly available \texttt{21cmFAST} code. The U-Net code is available on the author’s GitHub page: \url{https://github.com/samgagnon/wedge-unet}. The code used to generate instrumental noise realizations is available upon request.



\bibliographystyle{mnras}
\bibliography{biblio} 




\appendix

\section{Erratum}

\noindent This is a correction to Samuel Gagnon-Hartman, Yue Cui, Adrian Liu, Siamak Ravanbakhsh, Recovering the wedge modes lost to 21-cm foregrounds, \emph{Monthly Notices of the Royal Astronomical Society,} Volume 504, Issue 4, July 2021, Pages 4716–4729, \url{https://doi.org/10.1093/mnras/stab1158}. We will henceforth refer to the original manuscript as ``GH21".

During the drafting of a sequel paper \citep{2023arXiv230809740K} we found some undesirable properties in our splitting of the training and validation sets used to train our neural network. In general, the purpose of the validation set is to determine how well a network performs on data samples which it has not trained on. Our dataset took the form of a number of 21-cm brightness temperature coeval boxes, taken from a number of random seeds evaluated at multiple redshifts. While each box in the dataset is unique, many share their random seed in common and differ only in their redshift. This fact was not accounted for in GH21, resulting in random seeds being shared between members of the training and validation sets. In detail, 57 simulation boxes (generated with unique random seeds) were used in GH21. Of these, 38 were only present in training sets and therefore had no danger of cross-contamination. The remaining 19 boxes had some overlap in random seed. These are listed in Table~\ref{tab:randomseeds}. In principle, this cross contamination could allow the neural network to overfit the problem by memorizing the result from the training set and applying it to a member of the validation set with the same random seed.

Initially, the hope was that because no two boxes were exactly the same between the training and validation sets (in the sense of having \emph{both} the random seeds and the redshifts be the same; see Table~\ref{tab:randomseeds}), perhaps the separation in redshifts would be enough to avoid overfitting. Unfortunately, this was not the case, indicating that the problem was not just one that could occur in principle, but one that was being realized in practice. Remedying this issue by segregating random seeds between the validation and training sets, we found that the neural network---as originally proposed---cannot achieve results of the same quality presented in GH21.

Fortunately, we found that it is possible to restore the performance advertised in GH21 with just a few minor adjustments:
\begin{enumerate}
    \item An increase of the size of our datasets by a factor of $10$.
    \item Modifying the rate of spatial dropout to 0.3.
    \item Using a batch size of 3.\footnote{This precise batch size was selected due to computational limitations at the time the computations were performed. We find that varying the batch size from 2 to 6 has a minimal impact on the results.}
    \item Adding an extra regularization term to the cost function, such that the loss $D_{\rm loss}$ now reads
    \begin{equation}\label{eq:loss}
D_{\rm loss} = - \frac{2|X \cap Y|+\alpha}{|X|+|Y|+\alpha} +\alpha_{L_2} ||\boldsymbol{\omega}||^2_2,
\end{equation}
where $X$ is the array containing the ground truth data and $Y$ the array containing the prediction, with $\alpha =1$ and $\alpha_{L_2}=0.01$ as fixed hyperparameters. The vector $\boldsymbol{\omega} = (\omega_1,\omega_2,\dots)$ contains the model weights and $||\dots||^2_2$ denotes the $L_2$ norm. The first term in Equation~\eqref{eq:loss} is the same as it was in GH21; the second term is the additional regularization term, as suggested in, e.g., \citet{Goodfellow-et-al-2016} as a reasonably common method for combating overfitting.
\end{enumerate}

\begin{table}
    \centering
    \begin{tabular}{lll}
    \hline
    \textbf{Random seed} & \textbf{Redshift of boxes} & \textbf{Redshift of boxes}     \\ 
& \textbf{in training set} & \textbf{in validation set}     \\ \hline
    50                     & 8.00     & 8.10    \\ 
    400                     & 8.10       & 8.00     \\ 
    500                     & 8.50       & 8.10, 8.20, 8.30, 8.40     \\ 
    550                     & 8.20       & 8.30, 8.40, 8.50    \\ 
    850                     & 8.30       & 8.20, 8.40, 8.50    \\ 
    950                     & 7.85       & 8.00    \\ 
    \end{tabular}
    \caption{The random seeds and redshifts of the $19$ boxes that appeared in both training and validation sets of GH21. (Not shown are the remaining $38$ boxes that appeared only in the training set.) Although some random seeds are in common between training and validation sets, there are no cases where both the random seed and the redshift are the same. Nonetheless, the cross contamination was enough to affect the conclusions of GH21.}
    \label{tab:randomseeds}
\end{table}

\begin{figure}
    \centering
    \includegraphics[width=3 in]{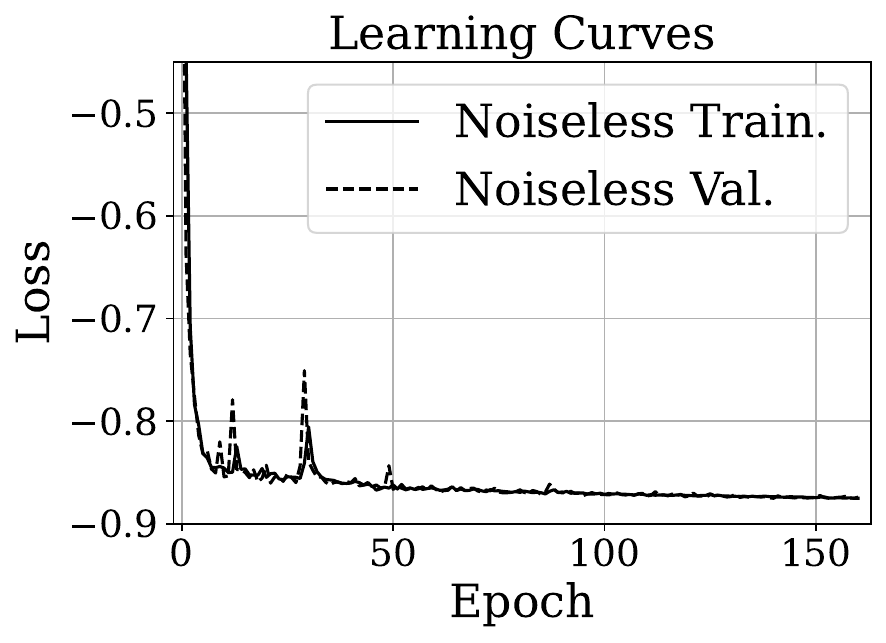}
    \caption{The binary dice loss computed at each epoch of training for the noiseless model. 
    A solid line is used for training loss, while the dashed line
    is used for validation loss. The curves are indicative of learning without
    significant over-fitting. In this re-analysis, we the trained the network on 
    data only affected by the wedge, and not by other instrumental effects, 
    hence we call the dataset ``noiseless".}
    \label{fig:loss}
\end{figure}
\begin{table}
    \centering
    \begin{tabular}{lllll}
    \hline
    \textbf{Neutral Fraction} & \textbf{Accuracy} & \textbf{Precision} & \textbf{Recall} & \textbf{IoU}   \\ \hline
    0.270                     & \underline{0.914}       & \underline{0.962}        & \underline{0.918}     & \underline{0.813}    \\
    0.351                     & 0.893             & 0.948              & 0.883           & 0.795          \\
    0.432                     & 0.871             & 0.934              & 0.831           & 0.770          \\
    0.510                     & 0.858             & 0.919              & 0.778           & 0.749          \\
    0.583                     & 0.845             & 0.902              & 0.706           & 0.718          \\
    0.645                     & 0.839             & 0.887              & 0.626           & 0.686          \\
    0.698                     & \textit{0.834}    & 0.870              & 0.531           & 0.647          \\
    0.745                     & 0.841             & \textit{0.830}     & \textit{0.472}  & \textit{0.625}
    \end{tabular}
    \caption{The tabulated statistics for the predictions made by the network on
    the validation data suite. The highest score in each column is underlined,
    while the lowest is italicized.}
    \label{tab:statistics}
\end{table}

With the above changes to our neural network, we are able to achieve comparable results as before. The training curve for the new network is shown in Figure \ref{fig:loss}, showing good training characteristics. New performance statistics are shown in Table \ref{tab:statistics} (analgous to Table 2 in GH21). Over a wider range of redshifts than in GH21, we find comparable accuracy and precision to be comparable (if a little worse) than before, while recall and IoU are consistently superior throughout. Sample network predictions are shown in Figure \ref{fig:pred-noiseless} (analogous to Figure 8 in GH21). Note that the training sets used for our present computations cover the same range of redshifts used in GH21; said differently, no changes other than those enumerated above were necessary.

GH21 presented three models, one trained on data that includes the ``wedge" foreground effect alone (with no instrumental noise), as well as a model each for data affected by the wedge plus the instrumental effects of HERA or the SKA. Here we have presented positive results for the noiseless case, where the reconstruction of ionized bubbles is successful. Unfortunately, we find that for the HERA case, this is no longer the case (although it is of course possible that a further-improved algorithm could restore the previous performance). For the SKA, a more expansive examination is presented in \citep{2023arXiv230809740K}, where the performance is comparable to before with no further changes to the algorithm beyond what is described here.

In conclusion, while our original treatment of the data in GH21 was flawed, in this Correction we show that our approach of using a U-Net to recover lost Fourier modes is still valid, and thus the main conceptual message of GH21 still stands.

\begin{figure*}
    \centering
    \includegraphics[width=5.8 in]{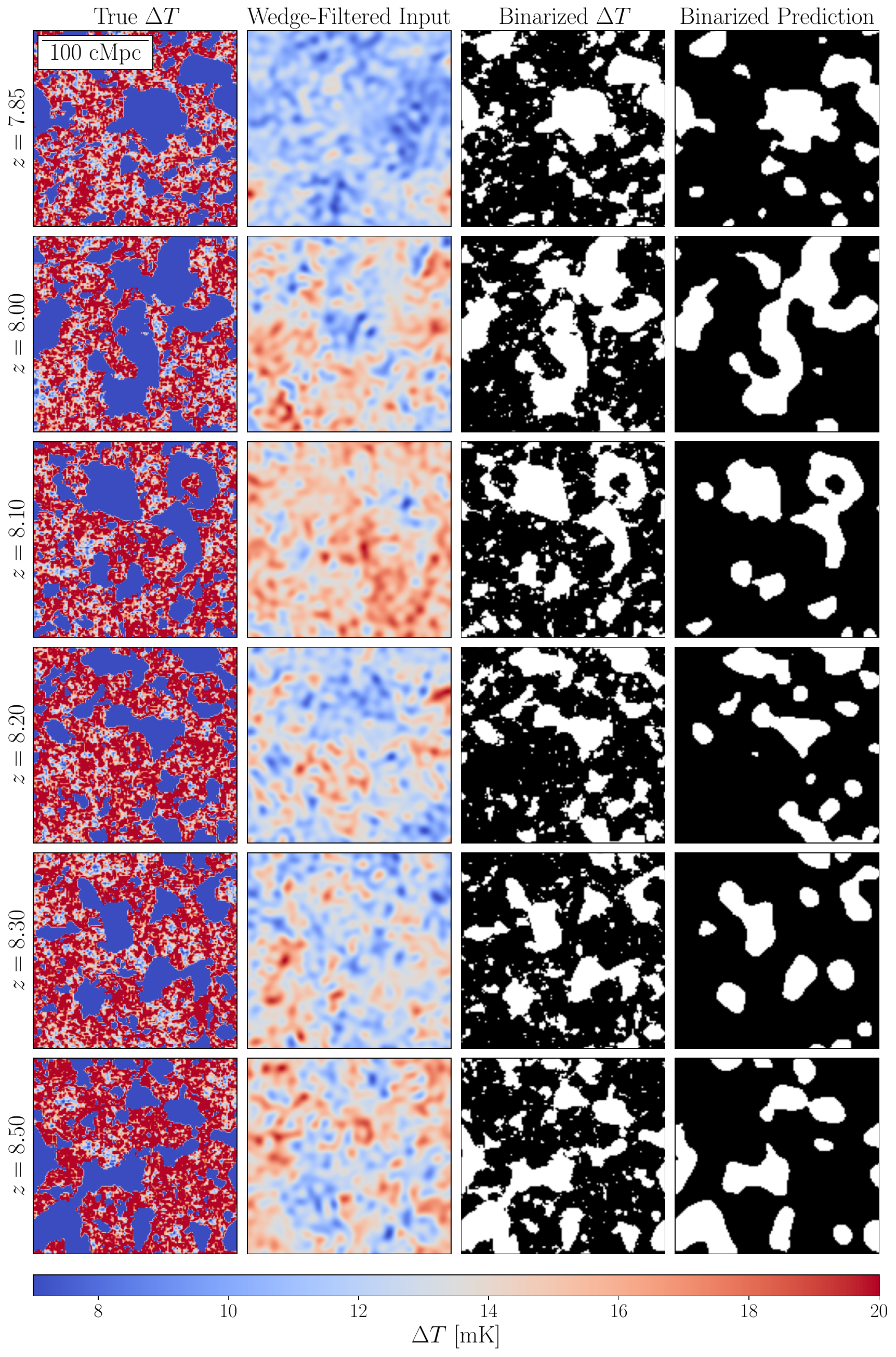}
    \caption{Sample network predictions on the noiseless data suite. The first column shows a transverse cross-section of the true brightness temperature field, while the second column shows the same field after excising Fourier modes lying within the foreground wedge. The third column is a binarized version of the first column and serves as the ground truth for our neural network. The fourth column shows the predicted ionization maps from our network. Visually, it is clear that our network is able to recover ionized bubbles from wedge-filtered 21-cm maps.}
    \label{fig:pred-noiseless}
\end{figure*}




\section*{Author Contributions}

Gagnon-Hartman, Cui, Liu, and Ravanbakhsh were the original authors of GH21. Gagnon-Hartman and Cui were responsible for most of the code development and execution of the project, based on Ravanbakhsh's ideas for the neural network architecture. The idea for the project originated from Liu, who also provided most of the guidance for interpreting the results. The bulk of the GH21 manuscript was written by Gagnon-Hartman and Liu, with editorial input from the rest of the co-authors.

Kennedy appears as a new co-author on this correction. He found the error in GH21, proposed the fixes described here, implemented them, and generated all the results. His contributions are limited to the noiseless amended model shown in this correction (although noisy extensions involving other effects are explored in a follow-up work; \citealt{2023arXiv230809740K}).

\bsp	
\label{lastpage}
\end{document}